\documentclass[%
 reprint,
superscriptaddress,
 amsmath,amssymb,
 aps,
prl,
floatfix,
]{revtex4-2}

\usepackage{graphicx}% Include figure files
\usepackage{dcolumn}% Align table columns on decimal point
\usepackage{bm}% bold math
\pdfoutput=1
\usepackage{blindtext}
\usepackage{amsfonts}
\usepackage{amsmath}
\usepackage{amssymb}
\usepackage{color}
\usepackage{epstopdf}

\usepackage[capitalize]{cleveref}
\DeclareSymbolFont{myletters}{OML}{ztmcm}{m}{it}
\DeclareMathSymbol{\uplambda}{\mathord}{myletters}{"15}
\begin{document}

\preprint{APS/123-QED}

\title{Topological Superconductivity in Rashba Spin-Orbital Coupling Suppressed Monolayer $\beta$-Bi$_{2}$Pd}% Force line breaks with \\

\author{Xin-Hai Tu}
\affiliation{Institute of High Energy Physics, Chinese Academy of Sciences (CAS), Beijing 100049, China}
\affiliation{University of Chinese Academy of Sciences, Beijing 100039, China}
\affiliation{Spallation Neutron Source Science Center, Dongguan 523803, China}
\author{Peng-Fei Liu}
\affiliation{Institute of High Energy Physics, Chinese Academy of Sciences (CAS), Beijing 100049, China}
\affiliation{Spallation Neutron Source Science Center, Dongguan 523803, China}
\author{Wen Yin}
\affiliation{Institute of High Energy Physics, Chinese Academy of Sciences (CAS), Beijing 100049, China}
\affiliation{University of Chinese Academy of Sciences, Beijing 100039, China}
\affiliation{Spallation Neutron Source Science Center, Dongguan 523803, China}
\author{Jun-Rong Zhang}
\affiliation{Institute of High Energy Physics, Chinese Academy of Sciences (CAS), Beijing 100049, China}
\affiliation{University of Chinese Academy of Sciences, Beijing 100039, China}
\affiliation{Spallation Neutron Source Science Center, Dongguan 523803, China}
\author{Ping Zhang}
\affiliation{School of Physics and Physical Engineering, Qufu Normal University, Qufu 273165, China}
\affiliation{Institute of Applied Physics and Computational Mathematics, Beijing 100088, China}
\author{Bao-Tian Wang}
\thanks{Author to whom correspondence should be addressed. E-mail: wangbt@ihep.ac.cn }
\affiliation{Institute of High Energy Physics, Chinese Academy of Sciences (CAS), Beijing 100049, China}
\affiliation{University of Chinese Academy of Sciences, Beijing 100039, China}
\affiliation{Spallation Neutron Source Science Center, Dongguan 523803, China}
\affiliation{Collaborative Innovation Center of Extreme Optics, Shanxi University, Taiyuan, Shanxi 030006, China}

\date{\today}% It is always \today, today,
\begin{abstract}
The weak interlayer Van Der Waals material $\beta$-Bi$_{\textrm2}$Pd has recently been established as a strong topological superconductor candidate with unconventional spin-triplet pairing and Majorana zero modes at vortices. In this letter, we study the topological characters and the superconducting pairing, which are still obscure in monolayer $\beta$-Bi$_{\textrm2}$Pd, in light of our effective theoretical model. We find that the non-Rashba spin-orbital coupling plays a critical role in realizing and tuning various novel topological natures. In particular, the spin-triplet p-wave superconducting pairing with Majorana zero mode is revealed in monolayer $\beta$-Bi$_{\textrm2}$Pd. Our studies deepen the understanding of topology and superconductivity in monolayer $\beta$-Bi$_{\textrm2}$Pd and indicate it is a promising platform for achieving low-dimentional topological superconductivity.
\end{abstract}
\maketitle
\paragraph*{Introduction}Since Majorana Fermions are their own antiparticles and obey non-Abelian braiding statistics \cite{Majorana1937,Moore1991}, they possess potential applications for fault-tolerant topological quantum computation \cite{Wilczek2009,Beenakker2013,Kitaev2003}. Topological superconductor (TSC) \cite{PhysRevLett.102.187001,Bernevig2013,PhysRevB.61.10267} with unconventional pairing symmetry is a natural platform for realizing topologically protected gapless boundary states which is essentially Andreev bound state hosting Majorana fermions \cite{Sato2017,RevModPhys.83.1057}. The p-wave superconductor Sr$_{\textrm2}$RuO$_{4}$ \cite{Rice_1995,Ishida1998,Kallin_2012} is a chiral TSC associated with spontaneous time-reversal (TR) symmetry breaking. Cu-doped Bi$_{\textrm2}$Se$_{\textrm3}$ \cite{PhysRevLett.104.057001,PhysRevLett.106.127004}, an odd-parity superconductor \cite{PhysRevLett.107.217001,PhysRevLett.108.057001}, is judged as the TR invariant TSC from the Fermi surface of normal state that encloses an odd number of time-reversal invariant momenta in the first Brillouin zone (BZ) \cite{PhysRevB.79.214526,PhysRevB.81.220504,PhysRevLett.105.097001}. In addition, a feasible project is proposed \cite{SATO2003126} to support non-Abelian anyon excitations, which is realized in two-dimensional (2D) Dirac Fermions, in s-wave superconductors. Subsequently, it is explicitly demonstrated by both theory \cite{PhysRevLett.100.096407} and experiment \cite{Zhang182,Yuan2019,Guan2016} that the Dirac-type surface state from a topological insulator couplings to a s-wave superconductor resembles a spinless p-wave superconductor where Majorana zero modes (MZMs) are realized at vortices.

Recently, a promising TSC candidate $\beta$-Bi$_{\textrm2}$Pd has attracted much attention on its topologically protected surface state \cite{Sakano2015,PhysRevB.100.161109,Wang2017} and fully-gapped anisotropic s-wave superconductivity \cite{PhysRevB.95.014512,PhysRevB.93.220504}. The angle-resolved photoemission
spectroscopy measurements on $\beta$-Bi$_{\textrm2}$Pd thin films show an anomalously large superconducting
gap of topological surface state \cite{GUAN20191215}. Furthermore, a signature of MZMs at vortices was observed via cryogenic scanning tunneling microscopy \cite{Lv2017}. Much interest is sparked by the observation of half-quantum magnetic flux quantization indicating an unconventional superconductor with the spin-triplet pairing symmetry \cite{Li2019}. Besides the bulk structure, fertile ground in monolayer $\beta$-Bi$_{\textrm2}$Pd is still uncultivated and worth equal studying owing to its manipulatable property and abundant topological physics inside.

In this letter, we are motivated to understand the topological and superconducting properties in monolayer $\beta$-Bi$_{\textrm2}$Pd. The bulk $\beta$-Bi$_{\textrm2}$Pd is a layered structure with tetragonal centrosymmetric space group I4/mmm as shown in Fig. 1 (a) which is drawn by VESTA \cite{Momma2008}. Using molecular beam epitaxy method, the monolayer structure could be synthized in experimment \cite{Denisov2017}. Our optimized lattice constant is $\emph a$ $=$ 3.322 \AA{}, which is slightly smaller than the experimental value 3.4 \AA{} of the $\beta$-Bi$_{\textrm2}$Pd films \cite{Lv2017}. Based on the reduced three-band tight-binding model (TBM), the complete process of topological phase transition is displayed by tuning spin-orbital coupling (SOC). Here, the Rashba SOC is suppressed. On the one hand, we find that the high-order quadratic Dirac point (DP) protected by C$_{\textrm4}$ rotation symmetry appears at the M point when ignoring the effect of SOC. On the other hand, including SOC, quantum spin Hall phase with nontrivial topological edge states (TESs) begin to emerge if the strength of SOC is less than 6.39 eV. The nontrival phase is characterized by the spin Chern number (SCN) $\emph C_{\emph s}=$ 1. In addition, we study pairing symmetry around the M point as well as the corresponding linearized gap equations, since monolayer $\beta$-Bi$_{\textrm2}$Pd is a superconductor according to first-principles calculations \cite{PhysRevB.102.155406}. Specifically, we find that monolayer $\beta$-Bi$_{\textrm2}$Pd is a Dirac superconductor with p-wave superconducting pairing when excluding SOC. Including SOC, it becomes a TSC of symmetry class D with MZMs at the boundary. The details of computation is shown in supplementary information (SI).
\begin{figure}
	\centering
	\includegraphics[width=7cm]{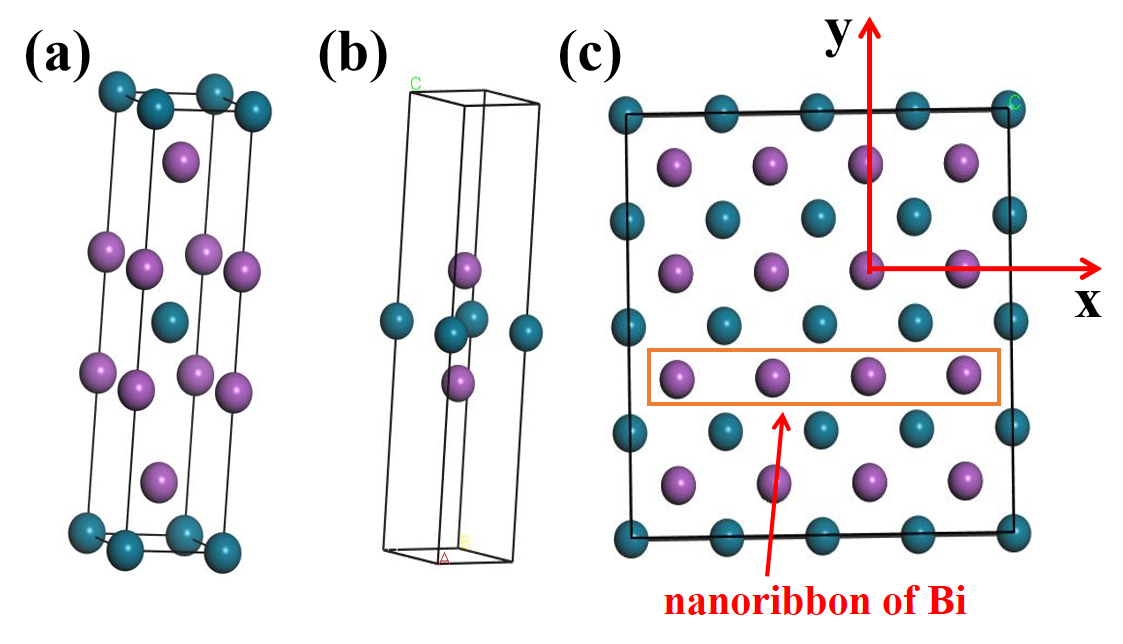}
	\caption{The bulk (a) and monolayer structure (b) of $\beta$-Bi$_{\textrm2}$Pd. The purple balls are Bi atoms, and navy blue balls are Pd atoms. (c) The top view of monolayer $\beta$-Bi$_{\textrm2}$Pd.}\label{fig1}
\end{figure}
\paragraph*{Basis and Model Hamiltonian}In order to get a physics picture of monolayer $\beta$-Bi$_{\textrm2}$Pd, we begin to choose suitable atomic orbitals. We find that the \emph{p}-orbitals of Bi atoms contribute mainly to the band structure around the Fermi level through the whole BZ, while the \emph{d}-orbitals of Pd atoms are away from \emph{p}-orbitals and only dominant on the valence bands from -5 to -2 eV, as shown in Fig. 2 (a). Meanwhile, the bands below -6 eV are disentangled with the above. Thus, we fit a eleven-band TBM by Wannier90 package \cite{Pizzi2020} first, which reproduce the bands around the Fermi level [red lines in Fig. 2 (b)]. Then, to simplify the problem, we treat the influence of Pd atoms as the perturbation and get a six-band TBM upto the fourth-nearest-neighbor hopping by downfolding technique \cite{doi:10.1063/1.1748067,PhysRevB.89.085130,2021leng} (see SI). Since the existence of inversion symmetry, it is natural to set up bonding and anti-bonding states with definite parity \cite{PhysRevB.82.045122} for Bi atoms as follow:
\begin{equation}
\begin{array}{cccc}
|\textrm{Bi}_{\emph x,\emph y,\emph z}^{\pm}\rangle=\frac{1}{\sqrt{2}}(|\textrm{Bi}_{\textrm1;\emph x,\emph y,\emph z}\rangle\mp|\textrm{Bi}_{\textrm2;\emph x,\emph y,\emph z}\rangle),
\end{array}
\end{equation}
where the superscript represents the parity. Eventually, we get a block diagonal six-band TBM (shown in SI). The fitting parameters are shown in Table S1. One can see that the contribution of Pd atoms does not introduce any additional hopping terms. In other words, in our case, we could equivalently only take Bi atoms into account to construct TBM and ignore the influence from the ineffective Pd atoms. One of the main reasons is that Bi and Pd atoms have the same site symmetry (4mm). 
\begin{figure}
	\centering
	\includegraphics[width=8.2cm]{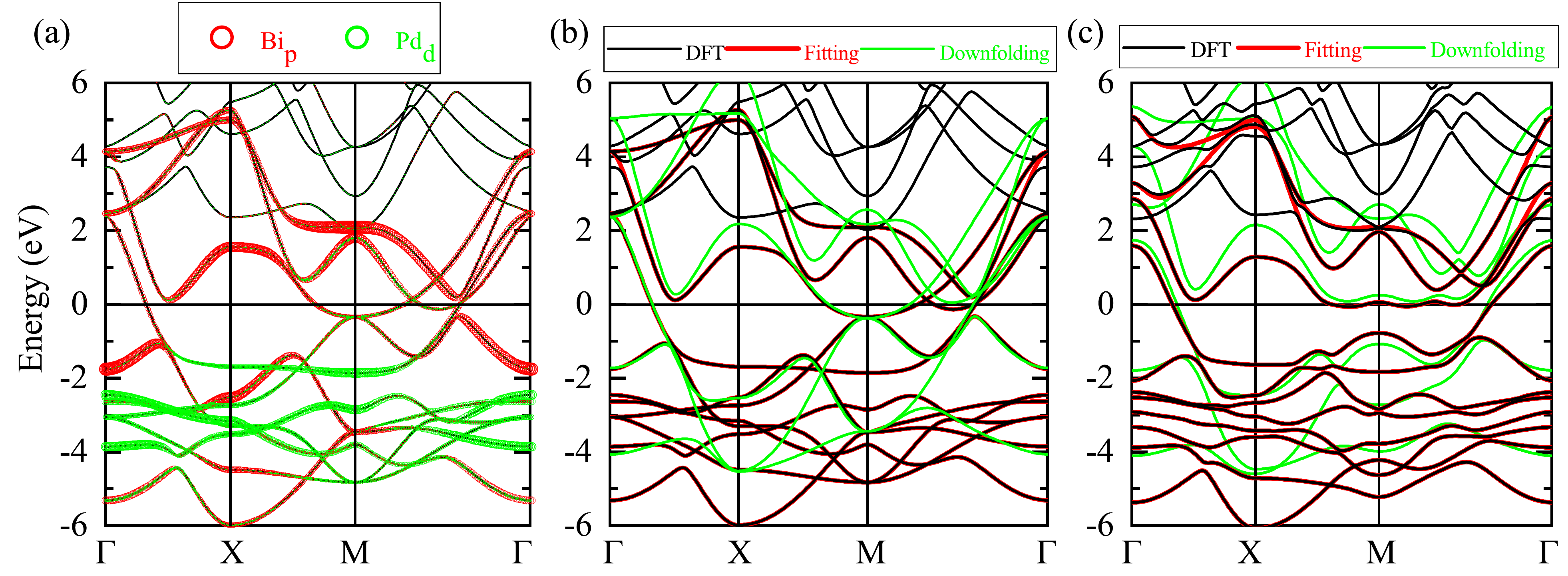}
	\caption{The band structure of monolayer $\beta$-Bi$_{\textrm2}$Pd without (a) (b) and with (c) SOC. The contribution of p$_{\emph z}$ and p$_{\emph x,\emph y}$ orbitals of Bi atoms are signed by red and green balls. The red lines represent the fitting eleven bands drawn according to wannier90. The green lines represent downfolding six bands.}\label{fig1}
\end{figure}
\paragraph*{Spin-orbital Coupling}
Owing to the heavy halogen family Bi, we consider the SOC effect in our model. Since potential field is the largest near the atomic nuclei, SOC is normally accurately approximated by a local atomic contribution of the form
\begin{equation}
\emph H_{\emph{SOC}}=\frac{\emph{dV}}{\emph{dr}}\frac{(\textbf{r}_{\emph i}\times \textbf{p}_{\emph i})\cdot \textbf{S}_{\emph i}}{\textrm2\emph r(\emph{mc})^{\textrm2}}=\textrm2\lambda \textbf{L}_{\emph i} \cdot \textbf{S}_{\emph i},
\end{equation}
where $\lambda=$ 0.578 eV and denotes the strength of SOC, $\textbf{L}_{\emph i}$ and $\textbf{S}_{\emph i}$ are orbital and spin angular momentum on site $\emph i$ of electron, respectively. Since the coincidence of TR and inversion symmetry, the Kramer degeneracy ensures the double degeneracy of each band. Therefore, it is no accident that SOC is nothing but splits the DP at the M point and causes a full band gap as shown in Fig. 2(c). Here, we ignore the SOC effect of Pd atoms which is much smaller than Bi ($\lambda_{\text{Pd}}=$ 0.087 eV). The in-plane SOC effect (Rashba SOC) coupling bands between bonding and anti-bonding states is excluded from consideration as well. Experimentally, it could be tunable by the electric field along z direction. So, we get the full Hamiltonian with SOC (see SI). In the case, the Hamiltonian is still block diagonal, which means that the spin $z$ component is not mixed and hence is still a good quantum number.
Next, we focus on $\emph H_{\emph S\textrm1}$ that we are interested and won’t go into details about $\emph H_{\emph S\textrm2}$ since $\emph H_{\emph S\textrm1}$ is independence of $\emph H_{\emph S\textrm2}$.
\paragraph*{High-Order Dirac Point}
Without SOC, there is a distinct DP \cite{PhysRevMaterials.3.054202,PhysRevB.88.125427,PhysRevB.98.081115} at the M point which is protected by C$_{\textrm4}$ rotation symmetry and is robust against any perturbations unless destroying the crystal symmetry. To illustrate the topological properties of the DP, we drop off the p$_{\emph z}$ orbital and reduce the TBM $\emph H_{\emph S\textrm 1}$ into the 2$\times$2 continuum model around the M point using perturbation theory, since the p$_{\emph z}$ orbital is far away from the other two orbitals \cite{Sun2011}.
In this case, the kernel of the Hamiltonian is still grasped from such virtual process in which a electron jumps from the $\emph p_{\emph x,y}$ to the $\emph p_{\emph z}$ orbital and then back to $\emph p_{\emph x,y}$ orbitals. Thus, the continuum Hamiltonian expanding at the M point reads (keep to the lowest terms of $\textbf k$)
\begin{equation}
\emph H_{\emph M}=\emph m_{\textrm3}\emph k_{x}\emph k_{y}\sigma_{\emph x}+\lambda\sigma_{\emph y}+\frac{\emph m_{\textrm1}-\emph m_{\textrm2}}{\textrm2}(\emph k_{\emph x}^{\textrm2}-\emph k_{\emph y}^{\textrm2})\sigma_{\emph z},
\end{equation}
where $\sigma$ is the Pauli matrix on the orbital space. 
Next, we define a 2D planar vector $\textbf d$ (see SI), which has a vortex structure at the DP ($\lambda$ = 0) from Fig. S2. This vortex is described by the winding number
\begin{equation}
\emph W=\oint_{\mathcal{C}} \frac{\emph d \textbf k}{\textrm 2 \pi} \cdot\left(\frac{\emph d_{\emph x}}{|\textbf d|} \nabla\frac{\emph d_{\emph y}}{|\textbf d|}-\frac{\emph d_{\emph y}}{|\textbf d|}\nabla\frac{\emph d_{\emph x}}{|\textbf d|}\right),
\end{equation}
which is 2 in our present case. Thus, the DP is a high-order DP with quadratic dispersions \cite{PhysRevB.101.205134,2009.12036}. One can find more details about the continuum model in SI.
\paragraph*{Topological phase transition}
By adiabatic continuity, as long as the Hamiltonian is gapped, it remains in the same topological phase (TP) unless it encounters gapless points which are TP transition points (TPTPs). The DP we mentioned before is exactly the TPTP connecting two different TPs. In fact, with the change of $\lambda$, the conduction and  valence bands also close at another critical point $\Gamma$ [Fig. S1 (b)] when $\lambda_{\textrm0}$ = 6.39 eV. Then we can write down the continuum model at the $\Gamma$ right now (upto the leading order)
\begin{equation}
\emph H_{\Gamma}=\gamma_{\textrm0}\sigma_{\textrm0}-\gamma_{\textrm1}\emph k_{\emph y}\sigma_{\emph x}+\gamma_{\textrm1}\emph k_{\emph x}\sigma_{\emph y}+\gamma_{\textrm2}\sigma_{\emph z}.
\end{equation}
Next, we show different topological phases as $\lambda$ goes from large to small \cite{Bernevig2013}. First of all, when $\lambda$ is extremely large upto infinity ($+\infty$), the Hamiltonian is toplogically equaivalent to the atomic limit, due to the flat and $\textbf k$-independent band structures. However, things start to change when we go through the first critical point $\Gamma$. The difference of SCN between $\lambda>\lambda_{\textrm0}$ and 0 $<\lambda<\lambda_{\textrm0}$ is -1, so SCN of the phase is nontrivial and given by $\emph C_{\emph s}$ $=$ -1. If we decrease $\lambda$ to zero, we reach another critical point M. The change of SCN is 2, thus the new phase has a nontrivial SCN $\emph C_{\emph s}$ $=$ 1. Finally, the Hamiltonian is back to the trivial state when $\lambda<-\lambda_{\textrm0}$ which is the same as the case $\lambda>\lambda_{\textrm0}$. SCN is calculated by integral Berry curvature $\emph F(\textbf{k})=\textrm i\nabla \emph u(\textbf k)\times\nabla \emph u(\textbf k)$ over the whole BZ as follow \cite{Chen2017,PhysRevLett.97.036808}
\begin{equation}
\emph C_{\emph s}=\frac{\textrm1}{\textrm2 \pi} \int \emph d^{\textrm2} \emph k \emph F(\textbf{k}),
\end{equation}
where $\emph u(\textbf k)$ is the wavefunction. Around the TPTPs $\Gamma$ and M, SCN is given by $\emph C_{\emph s}=\frac{\textrm1}{\textrm2}\textrm{sgn}(\eta)$ and $\emph C_{\emph s}=\textrm{sgn}[\lambda\emph m_{\textrm3}(\emph m_{\textrm1}-\emph m_{\textrm2})]$, respectively, where $\eta=\pm$ 1. Meanwhile, the existence of inversion symmetry helps us to describe the nontrivial TP by $\mathcal{Z}_{\textrm2}$ index via parity method \cite{PhysRevLett.98.106803,fu2007topological} as shown in Table S2. There is a distinct band inversion along $\Gamma$-X line that results in nontrivial $\mathcal{Z}_{\textrm2}$ = 1.
\begin{figure}
	\centering
	\includegraphics[width=7cm]{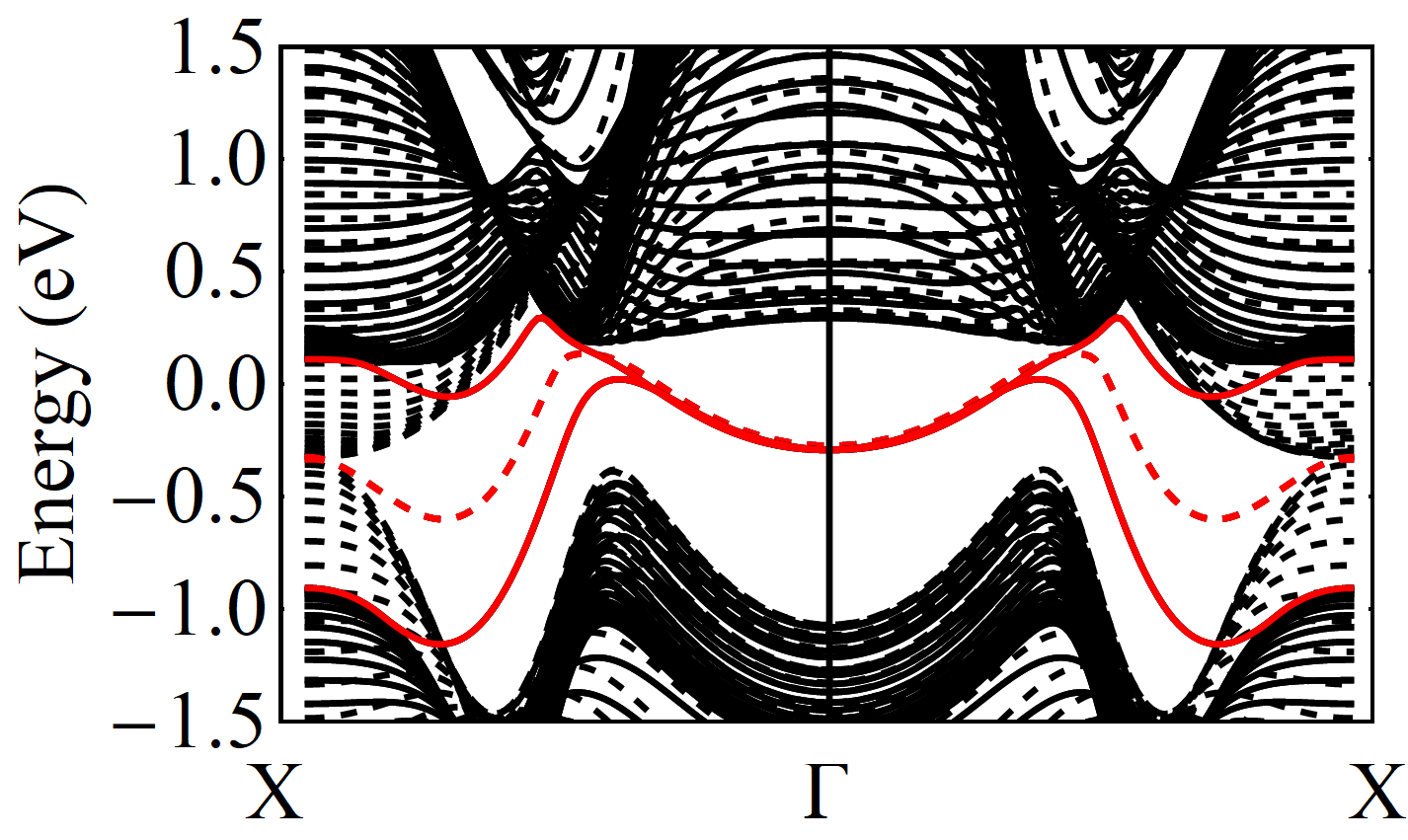}
	\caption{The projected band structures along [01] direction with (solid lines) and without (dashed lines) SOC. TESs are marked with red lines.}\label{fig1}
\end{figure}
\paragraph*{Topological Edge State}Besides SCN, another remarkable feature of a nontrivial topological phase is toplogical boundary state. Considering an infinite nanoribbon of monolayer $\beta$-Bi$_{\textrm2}$Pd where $\emph y$ direction is limited (Fig. 1), in the case, the momentum component $\emph k_{\emph y}$ is not a good quantum number yet. In Fig. 3, we plot the projected edge band structure of nontrivial TP for $\emph H_{\emph S\textrm 1}$. Excluding SOC, we can see a TES connects two projected DPs as in graphene model \cite{Ezawa2015}. Once turning on SOC, the band structure in bulk will be divided and then forms a full band gap. However, the TESs close the band gap at the $\Gamma$ point (solid red lines). The TESs corresponding to quantum spin Hall state \cite{Qian2014,Ezawa2015} that two electrons with opposite spin travel toward adverse directions at the boundary since the system preserves the TR symmetry.
\begin{table}[ptb]
	\caption{Classification of all on-site and nearest neighbor pairing potentials according to the representations of D$_{\textrm4\emph h}$ point group.}
	\begin{ruledtabular}
		\begin{tabular}{c|cccccccccccccc}
			Form factors&Representation &Matrix form &Spin\\
			\hline
			A$_{1g}$/B$_{1g}$,$\cos(\emph{kx})\pm\cos(\emph{ky})$&$\emph A_{\textrm1\emph g}$,$\Delta_{\alpha\textrm1}$& $\sigma_{\textrm0}s_{\textrm0}$&Singlet\\
			&$\emph A_{\textrm1\emph g}\Delta_{\alpha\textrm2}$&$\sigma_{\emph y}\emph s_{\emph z}$&Triplet\\
			&$\emph B_{\textrm1\emph g}$,$\Delta_{\alpha\textrm3}$&$\sigma_{\emph z}s_{\textrm0}$&Singlet\\
			&$\emph B_{\textrm2\emph g}$,$\Delta_{\alpha\textrm4}$&$\sigma_{\emph x}s_{\textrm0}$&Singlet\\
			&$\emph E_{\emph g}$,($\Delta_{\alpha\textrm5}$,$\Delta_{\alpha\textrm6}$)&($\sigma_{\emph y}s_{\emph x}$,$\sigma_{\emph y}s_{\emph y}$)&Triplet\\
			\hline
			E$_{u}$,$[\sin(\emph{kx}),\sin(\emph{ky})]$&$\emph E_{\emph g}$&($\sigma_{\textrm0}s_{\emph x}$,$\sigma_{\textrm0}s_{\emph y}$)&Triplet\\
			&$\emph E_{\emph g}$&($\sigma_{\emph z}s_{\emph x}$,$\sigma_{\emph z}s_{\emph y}$)&Triplet\\
		\end{tabular}
	\end{ruledtabular}
\end{table}
To show the validity of our model further, in Fig. S3, we show the Bi terminated TESs calculated by WannierTools package \cite{2018wu} as a control, whose main physical features are captured in our effective six-band model.
\paragraph*{Pairing Symmetry and Majorana Zero Mode}
Below the superconducting transition temperature (STT) $\emph T_{\emph c}=$ 1.95 K \cite{PhysRevB.102.155406}, Cooper pairs are formed from two electrons occupying orbitals of $|\textrm{Bi}_{\emph x,\emph y}^{\pm}\rangle$. According to U-V model \cite{Sato2017,PhysRevLett.105.097001}, we classify all possible on-site and nearest neighbor superconducting pairing potentials $\Delta_{\alpha\emph i}(\textbf k)=\Delta_{\text0}\emph f_{\alpha}(\textbf k)\Gamma_{i}$ under the constraint of point group D$_{\text4\emph h}$ in Table I, where U and V are intraorbital and interorbital interactions, respectively, $\alpha$ is pairing symmetry index, $\emph f_{\alpha}(\textbf k)$ is pairing form factor, and $\Gamma_{i}$ is the $\emph i^{\emph{th}}$ matrix form of irreducible representation of pairing potentials.  For p-wave superconducting states, we consider the states with TR and C$_{\text4}$ rotation symmetry. Thus, only two odd-parity pairings survive: $\Delta_{p\text1}(\textbf k)=\sin\emph{k}_{\emph x}\sigma_{\text0}\emph s_{\emph x}+\sin\emph{k}_{\emph y}\sigma_{\text0}\emph s_{\emph y}$ and $\Delta_{\emph p\text2}(\textbf k)=\sin\emph{k}_{\emph x}\sigma_{\emph z}\emph s_{\emph x}-\sin\emph{k}_{\emph y}\sigma_{\emph z}\emph s_{\emph y}$ (note that these components $\sin(\emph{k}_{\emph x}/\emph{k}_{\emph y})\sigma_{\text0/\emph z}\emph s_{\emph x}\pm\sin(\emph{k}_{\emph y}/\emph{k}_{\emph x})\sigma_{\text0/\emph z}\emph s_{\emph y}$ are
equivalent). Both of them are belong to the representation A$_{1u}$. On the Nambu basis \{$\emph C_{\textbf{k}\emph z,\uparrow}^{\dagger}$, $\emph C_{\textbf{k}\emph x,\uparrow}^{\dagger}$, $\emph C_{\textbf{k}\emph y,\uparrow}^{\dagger}$, $\emph C_{\textbf{k}\emph z,\downarrow}^{\dagger}$, $\emph C_{\textbf{k}\emph x,\downarrow}^{\dagger}$, $\emph C_{\textbf{k}\emph y,\downarrow}^{\dagger}$, $\emph C_{-\textbf{k}\emph z,\downarrow}$, $\emph C_{-\textbf{k}\emph x,\downarrow}$, $\emph C_{-\textbf{k}\emph y,\downarrow}$, $-\emph C_{-\textbf{k}\emph z,\uparrow}$, $-\emph C_{-\textbf{k}\emph x,\uparrow}$, $-\emph C_{-\textbf{k}\emph y,\uparrow}$\}, the Bogoliubov-de Gennes (BdG) Hamiltonian for $\emph H_{\emph S\textrm1}$ is given by
\begin{equation}
\begin{aligned}
\emph H_{\emph{BdG}\textrm1}=\left[\begin{array}{cc}
\emph H_{\emph S\textrm1}(\textbf{k}) & \Delta_{\alpha\emph i} \\
\Delta_{\alpha\emph i}^{\dagger} & -\emph s_{\emph y}\emph H_{\emph S\textrm1}^{*}(-\textbf{k})\emph s_{\emph y}
\end{array}\right].
\end{aligned}
\end{equation}
In Figs. S4-S7, we show the superconducting band structures for different pairing potentials. For s-wave pairing without SOC, the spin-singlet pairing $\Delta_{\emph s\textrm1}$ gives a full superconducting gap. Nevertheless, when monolayer $\beta$-Bi$_{\textrm2}$Pd enters $\Delta_{\emph s\textrm2}$ phase, it becomes a nodal loop superconductor where the loop surrounds the M point which is the same as $\Delta_{\emph s\textrm5}$ ($\Delta_{\emph s\textrm5}$ is equivalent to $\Delta_{\emph s\textrm6}$). For $\Delta_{\emph s\textrm3}$ ($\Delta_{\emph s\textrm4}$) pairing, the DP appears on the M-$\Gamma$ (X-M) line. Once turning on SOC, pairings $\Delta_{\emph s\textrm2}$ will open a superconducting band gap. If SOC strength is larger than the gap function strength $|\lambda|>|\Delta_{\textrm0}|$, $\Delta_{\emph s\textrm3/4/5}$ will be gapped as well. For s$^{*}$-wave, it is the same as s-wave. However, for d-wave pairing, due to the form factor of $\emph f_{d}(\textbf k)=\cos\emph k_{\emph x}-\cos\emph k_{\emph y}$, there are always DPs along M-$\Gamma$ line for all channels without SOC. With SOC, if $|\lambda|$ is larger enough, the superconducting gaps will appear. For p-wave, $\Delta_{\emph p\textrm1}$ give s a superconducting band gap but $\Delta_{\emph p\textrm2}$ has a Dirac point along M-$\Gamma$ line without SOC. With SOC, these two pairings are gapped.

The continuum model at the M point is illustrative of issues. Here, we take pairing potential $\Delta_{\emph s\textrm2}$ as an example. The superconducting eigenvalue is given in SI. One can see when $\lambda$ is zero, a Fermi loop around the M point is obtained by solving such $\textbf k$-equation $[\frac{(\emph m_{\textrm1}-\emph m_{\textrm2})}{\textrm2}(\emph k_{\emph x}^{\textrm2}-\emph k_{\emph y}^{\textrm2})]^{\textrm2}+(\emph m_{\textrm3}\emph k_{\emph x}\emph k_{\emph y})^{\textrm2}=\Delta_{\textrm0}^{\textrm2}$. The diagrammatic sketch is shown in Fig. S8 (a) in which the loop is actually a rounded square. More details of other pairings are shown in SI.

Next to determine the possible superconducting pairings, we evaluate the linearized gap equations in each pairing potential (see SI). In the strong SOC limit ($\lambda\approx\mu$), the possible channel in s-wave pairing is $\emph A_{\text1\emph g}$ which gives the STT by $\emph k_{\emph B} \emph T_{\emph c}=\frac{\textrm2 \emph e^{\gamma}}{\pi}\omega_{\emph D} \exp \left(-\frac{\textrm1}{\textrm2 \emph g(\textrm0) \emph V_{\emph{eff}}}\right)$. Here, $\mu$ is the chemical potential, $\omega_{\emph D}$ is the Debye frequency, $\gamma \approx$ $0.5772$ is the Euler constant and $\emph g(\textrm0)$ is the density of states at the Fermi level. The effective interaction $\emph V_{\emph{eff}}=U+V\frac{\lambda^{2}}{\mu^{2}}\approx U+V$. For non-on-site pairing, rather than performing an accurate calculation of STT \cite{PhysRevLett.101.206404,PhysRevX.1.011009,Hu2012}, we qualitative evaluate the relative relation of STT in each channel. The possible pairings for s$^{*}$-wave and d-wave are the same as s-wave. For p-wave pairing, $\Delta_{p\text1}$ has a vanishing STT and the possible superconducting ground state is $\Delta_{p\text2}$. In Fig. 4, we show TESs for these possible superconducting pairings in which $\Delta_{d/p}$ has nontrivial MZM. It is proved that two copies of the chiral spinless p-wave superconductor is a intrinsic TR invariant TSC \cite{PhysRevB.61.10267}. As shown in Fig. S9, even if SOC is excluding, MZM still survives in pairing $\Delta_{\emph p\text1}$. In view of the experimental observation of the spin-triplet p-wave superconducting pairing \cite{Li2019}, $\Delta_{p\text2}$ is considered to be a strong pairing candidate in monolayer $\beta$-Bi$_{\textrm2}$Pd. In the case, the BdG Hamiltonian commutes with $\emph z$ component of spin $[\emph H_{\emph{BdG1}},\emph S_{\emph z}]=$ 0, which implies that the Hamiltonian can be brought into block-diagonal form. While these two blocks are related by TR, each block is TR breaking \cite{PhysRevB.97.224507}. As a consequence, each block Hamiltonian belongs to symmetry class D and the corresponding topological invariant is $\mathcal Z$ in 2D according to ten symmetry classes of topological systems \cite{PhysRevB.78.195125}.
\begin{figure}
	\centering
	\includegraphics[width=8cm]{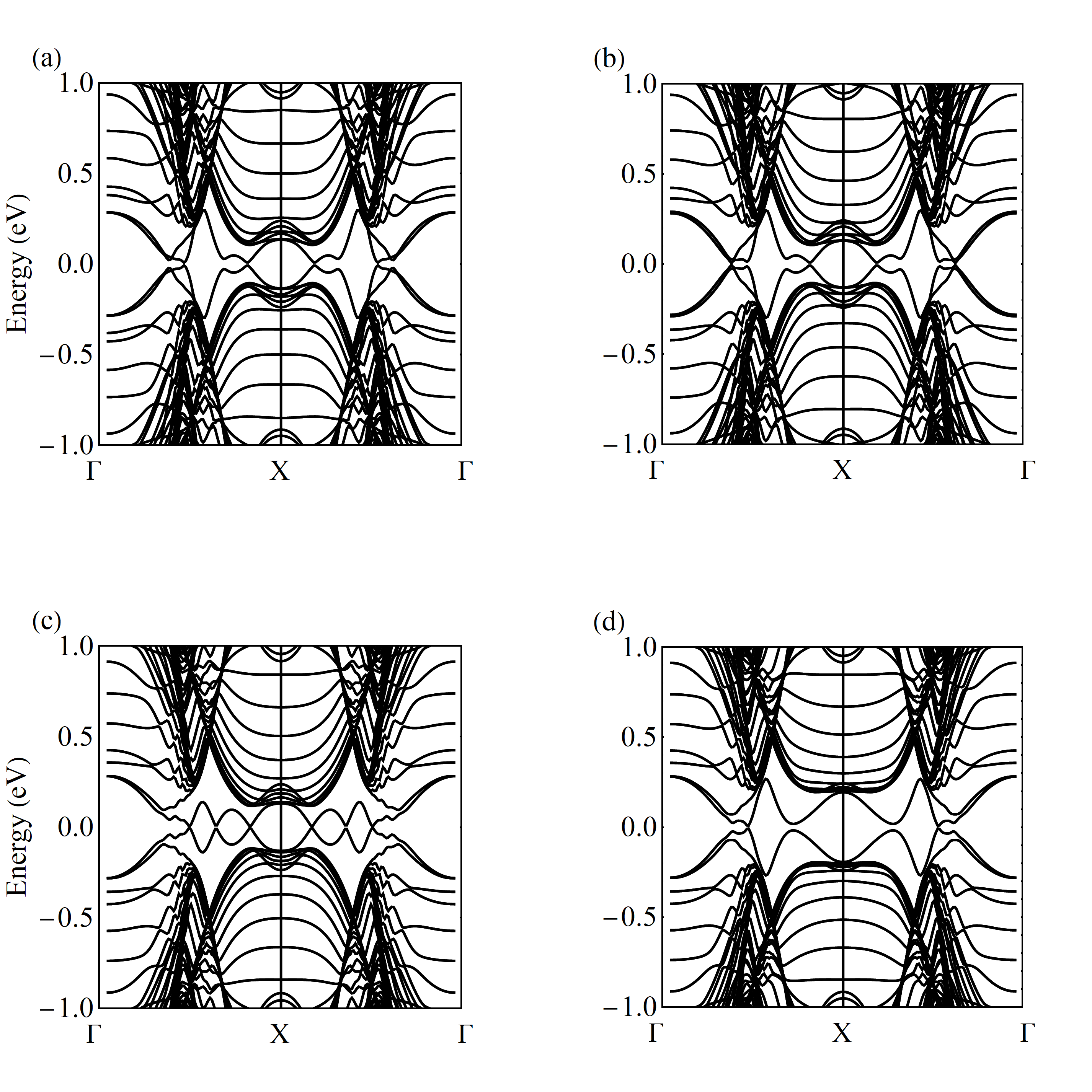}
	\caption{The Majorana Zero modes for different pairing potentials (a) $\Delta_{\emph d\textrm1}$, (b) $\Delta_{\emph d\textrm2}$, (c) $\Delta_{\emph p\textrm1}$, (d) $\Delta_{\emph p\textrm2}$. Here, we set $\lambda=0.578$ eV and $\Delta_{0}=0.3$ eV as examples.}\label{fig1}
\end{figure}
\paragraph*{Conclusion and Discussion}
In summary, we have presented a systematic study on topological and superconducting properties of monolayer $\beta$-Bi$_{\textrm2}$Pd. First, we show that the effect of Pd atoms is inessential to the interested physics. The effective TBM constructed from Bi atoms can capture the main physics. Second, the previous scheme shows that Rashba SOC with Zeeman field is a possible way to realize topological superconductivity \cite{PhysRevLett.103.020401,PhysRevB.82.134521}. Based on our model, however, we find that the non-Rashba SOC effect also plays a key role for realizing and tuning various exotic topological phenomena, such as high-order DP, quantum spin Hall state, Dirac superconducting state, nodal loop superconducting state and topological superconducting state. Finally, our results show that monolayer $\beta$-Bi$_{\textrm2}$Pd is a strong 2D TSC candidate of symmetry class D in the spin-triplet p-wave superconducting pairing channel. We hope this work could provide inspiration and guidance for further experimental and theoretical works in monolayer $\beta$-Bi$_{\textrm2}$Pd.
\paragraph*{Acknowledgements}
We thank Dr. Yun-Long Lian for helpful discussions. The authors gratefully acknowledge financial support
from and National Natural Science Foundation of China (Grant No. 12074381) and Science Challenge Project (Grant No.
TZ2016001). The authors also are thankful for the computational resources from the Supercomputer Centre of the China
Spallation Neutron Source.
\bibliographystyle{apsrev4-1}
\bibliography{bibl}
\end{document}

\preprint{APS/123-QED}

\title{Supplementary Information\\Topological Superconductivity in Rashba Spin-Orbital Coupling Suppressed Monolayer $\beta$-Bi$_{2}$Pd
}% Force line breaks with \\

\author{Xin-Hai Tu}
\affiliation{Institute of High Energy Physics, Chinese Academy of Sciences (CAS), Beijing 100049, China}
\affiliation{University of Chinese Academy of Sciences, Beijing 100039, China}
\affiliation{Spallation Neutron Source Science Center, Dongguan 523803, China}
\author{Peng-Fei Liu}
\affiliation{Institute of High Energy Physics, Chinese Academy of Sciences (CAS), Beijing 100049, China}
\affiliation{Spallation Neutron Source Science Center, Dongguan 523803, China}
\author{Wen Yin}
\affiliation{Institute of High Energy Physics, Chinese Academy of Sciences (CAS), Beijing 100049, China}
\affiliation{Spallation Neutron Source Science Center, Dongguan 523803, China}
\author{Jun-Rong Zhang}
\affiliation{Institute of High Energy Physics, Chinese Academy of Sciences (CAS), Beijing 100049, China}
\affiliation{Spallation Neutron Source Science Center, Dongguan 523803, China}
\author{Ping Zhang}
\affiliation{School of Physics and Physical Engineering, Qufu Normal University, Qufu 273165, China}
\affiliation{Institute of Applied Physics and Computational Mathematics, Beijing 100088, China}
\author{Bao-Tian Wang}
\thanks{Author to whom correspondence should be addressed. E-mail: wangbt@ihep.ac.cn }
\affiliation{Institute of High Energy Physics, Chinese Academy of Sciences (CAS), Beijing 100049, China}
\affiliation{University of Chinese Academy of Sciences, Beijing 100039, China}
\affiliation{Spallation Neutron Source Science Center, Dongguan 523803, China}
\affiliation{Collaborative Innovation Center of Extreme Optics, Shanxi University, Taiyuan, Shanxi 030006, China}
\maketitle

%\tableofcontents
\section{\label{sec:level3}Computational details}
Our calculations were performed within the QUANTUM-ESPRESSO package \cite{Giannozzi2009} based on density functional theory \cite{hohenberg1964inhomogeneous,kohn1965self} with ultrasoft pseudopotentials \cite{PhysRevB.41.7892,PhysRevLett.85.5122}. We used the Perdew-Burke-Ernzerhof type \cite{PhysRevLett.77.3865} of generalized gradient approximation  to evaluate the exchange-correlation functional. The kinetic-energy cutoff of the plane wave expansion is 60 Ry. A 15 \AA{} vacuum slab and a 32 $\times$ 32 $\times$ 1 $\textbf k$ mesh are adopted in 2D BZ. The internal atomic positions were fully relaxed with a threshold of 10 meV/\AA{} for the forces. The maximally localized Wannier functions for all \emph{p}-orbitals of Bi
and \emph{d}-orbitals of Pd are generated to construct the eleven-band TBM based on Wannier90 package \cite{Pizzi2020}. With the help of WannierTools codes \cite{2018wu}, we calculate the edge states in Fig. S3 (a) and (c) to illustrate the validity of our model.

\section{\label{sec:level3} Tight-binding model and Downfolding technique}
In this section, we introduce TB method first. The Bloch wave function can be written as follow
\begin{equation}
\left|\psi_{\alpha, \emph j}^{\textbf{k}}\right\rangle=\frac{1}{\sqrt{\emph N}}\sum_{\textbf{R}_{\emph n}}e^{\imath \textbf{k} \cdot \textbf{R}_{\emph n}} \left|\phi_{\textbf{R}_{\emph n}, \alpha, \emph j} \right\rangle,
\end{equation}
where $\left|\phi_{\textbf{R}_{\emph n}, \alpha, \emph j} \right\rangle$ represents the $\alpha^{th}$ atomic orbital of $\emph j^{\emph{th}}$ atom, $\textbf{R}_{\emph n}$ is the lattice vector, and N is the number of primitive cells. The matrix form of Hamiltonian is
\begin{equation}
\emph H_{\alpha \emph i, \beta \emph j}^{\textbf{k}} \equiv\left\langle\psi_{\alpha i}^{\textbf{k}}|\hat{\emph H}| \psi_{\beta j}^{\textbf{k}}\right\rangle=\sum_{\textbf{R}_{\emph n}} e^{\imath \textbf{k} \cdot \textbf{R}_{\emph n}}\emph E_{\alpha \beta}^{\emph{ij}}(\textbf{R}_{\emph n}),
\end{equation}
where
\begin{equation}
\emph E_{\alpha \beta}^{\emph i\emph j}(\textbf{R}_{\emph n})=\left\langle\phi_{\alpha \emph i}(\textbf{r})|\hat{\emph H}| \phi_{\beta j}(\textbf{r}-\textbf{R}_{\emph n})\right\rangle
\end{equation}
is the hopping integral between the atomic orbitals $\left|\phi_{\alpha \emph i}\right\rangle$ at
$\textbf{0}$ and $\left|\phi_{\beta \emph j}\right\rangle$ at lattice vector $\textbf{R}_{\emph n}$. Given $E_{\alpha \beta}^{\emph i\emph j}(\textbf{R}_{\emph n}),$ the hopping integrals to all neighboring sites can be generated by
\begin{equation}
\emph E^{\emph i\emph j}(\emph R\textbf{R}_{\emph n})=\emph D^{\emph i}\left(\emph R\right) \emph E^{\emph i\emph j}(\textbf{R}_{\emph n})\left[\emph D^{\emph j}\left(\emph R\right)\right]^{\dagger}
\end{equation}
where $\emph D^{\emph i}\left(\emph R\right)$ is the matrix of the $\emph i^{\emph{th}}$ irreducible representations of symmetry operations $\emph R$ in crystal point group and $E^{\emph{ij}}(\textbf{R}_{\emph n})$ is the matrix form of $\emph E_{\alpha \beta}^{\emph{ij}}(\textbf{R}_{\emph n})$.  Under the orbital basis of \{$|\textrm{Bi1}_{\emph z}\rangle$,$|\textrm{Bi1}_{\emph x}\rangle$,$|\textrm{Bi1}_{\emph y}\rangle$,$|\textrm{Bi2}_{\emph z}\rangle$,$|\textrm{Bi2}_{\emph x}\rangle$,$|\textrm{Bi2}_{\emph y}\rangle$,$|\textrm{Pd}_{\emph z^{\textrm2}}\rangle$,$|\textrm{Pd}_{\emph{zx}}\rangle$,$|\textrm{Pd}_{\emph{zy}}\rangle$,$|\textrm{Pd}_{\emph x^{\textrm2}-\emph y^{\textrm2}}\rangle$,$|\textrm{Pd}_{\emph{xy}}\rangle$\}, We construct a eleven-band TBM into
blocks by Wannier90 as follow
\begin{equation}
\begin{array}{cccc}
\emph H=\left[\begin{array}{lllllllll}
\emph H_{\text{Bi}} & A\\A^{\dagger}&\emph H_{\text{Pd}}
\end{array}\right].
\end{array}
\end{equation}
Since the energies of Pd atoms are far removed from the main orbitals of interest [Fig. 2 (a)], we could include their effects on the block $\emph H_{\text{Bi}}$ by Löwdin downfolding technique \cite{doi:10.1063/1.1748067,PhysRevB.89.085130,2021leng} which is in essence a perturbative method:
\begin{equation}
\emph H_{\text{0}}=\emph H_{\text{Bi}}+A(E-\emph H_{\text{Pd}})^{-1}A^{\dagger}.
\end{equation}
Eq. (6) can be solved by an iterative method which the lowest order yields the result
\begin{equation}
\emph H_{\text0}^{\emph{ij}}=\emph H_{\text{Bi}}^{\emph{ij}}+\sum_{k}\dfrac{A^{ik}A^{kj}}{E-\emph H_{\text{Pd}}^{\emph{kk}}},
\end{equation}
where i and j belong to the subspace $\emph H_{\text{Bi}}$ and k belongs to the subspace $\emph H_{\text{Pd}}$. The second term on the right hand side is from the second order perturbation, which describes the virtual hopping processes from the state $|\text{Bi}\rangle$ to $|\text{Pd}\rangle$ and to $|\text{Bi}\rangle$.
Here, we take E = -0.3 eV. Finally, by a unitary transformation $\emph{UH}_{\text0}\emph U^{\dagger}$,
\begin{equation}
\emph U=\dfrac{1}{\sqrt{2}}\left(\begin{array}{lllllll}
1& 0&0&1&0&0\\
0 &1&0&0&-1&0\\
0&0&1&0&0&-1\\
1&0&0&-1&0&0\\
0&1&0&0&1&0\\
0&0&1&0&0&1
\end{array}\right),
\end{equation}
we get the bonding and anti-bonding states of basis 1: \{$|\textrm{Bi}_{\emph z}^{-}\rangle,|\textrm{Bi}_{\emph x}^{+}\rangle,|\textrm{Bi}_{\emph y}^{+}\rangle$\} and basis 2: \{$|\textrm{Bi}_{\emph z}^{+}\rangle,|\textrm{Bi}_{\emph x}^{-}\rangle,|\textrm{Bi}_{\emph y}^{-}\rangle$\} and the corresponding six-band TBM upto the fourth-nearest-neighbor hopping is given as follow:
\begin{equation}
\begin{array}{cccc}
\emph H_{\textrm0}=\left[\begin{array}{lllllllll}
\emph H_{\textrm1} & 0\\0&\emph H_{\textrm2}
\end{array}\right]
=\left[\begin{array}{lllllllll}
\emph h_{\textrm1\textrm1} & \emph h_{\textrm1\textrm2} & \emph h_{\textrm1\textrm3}&&& \\
& \emph h_{\textrm2\textrm2} & \emph h_{\textrm2\textrm3}&&0& \\
& & \emph h_{\textrm3\textrm3}&&&\\
&&&\emph h_{\textrm4\textrm4}&\emph h_{\textrm4\textrm5}&\emph h_{\textrm4\textrm6}\\
&h.c.&&&\emph h_{\textrm5\textrm5}&\emph h_{\textrm5\textrm6}\\
&&&&&\emph h_{\textrm6\textrm6}
\end{array}\right],
\end{array}
\end{equation}
where
\begin{equation}
\begin{array}{cccc}
\emph h_{\textrm1\textrm1}=\epsilon_{\textrm1}+\textrm2\emph t_{\textrm1\textrm1}[\cos(\emph k_{\emph x}\emph a)+\cos(\emph k_{\emph y}\emph a)]+\textrm4\emph t_{\textrm1\textrm1}^{'}\cos(\emph k_{\emph x}\emph a)\cos(\emph k_{\emph y}\emph a)\\
+\textrm2\emph t_{\textrm1\textrm1}^{''}[\cos(\textrm2\emph k_{\emph x}\emph a)+\cos(\textrm2\emph k_{\emph y}\emph a)]+\textrm4\emph t_{\textrm1\textrm1}^{'''}[\cos(\textrm2\emph k_{\emph x}\emph a)\cos(\emph k_{\emph y}\emph a)+\cos(\emph k_{\emph x}\emph a)\cos(\textrm2\emph k_{\emph y}\emph a)],\\

\emph h_{\textrm1\textrm2}=\textrm2\textrm{i}\emph t_{\textrm{12}}\sin(\emph k_{\emph x}\emph a)+\textrm4\textrm{i}\emph t_{\textrm1\textrm2}^{'}\sin(\emph k_{\emph x}\emph a)\cos(\emph k_{\emph y}\emph a)\\
+\textrm2\textrm{i}\emph t_{\textrm{12}}^{''}\sin(\textrm2\emph k_{\emph x}\emph a)+\textrm4\textrm{i}\emph t_{\textrm1\textrm2}^{'''}\sin(\emph k_{\emph x}\emph a)\cos(\textrm2\emph k_{\emph y}\emph a),\\

\emph h_{\textrm1\textrm3}=\textrm2\textrm{i}\emph t_{\textrm{12}}\sin(\emph k_{\emph y}\emph a)+\textrm4\textrm{i}\emph t_{\textrm1\textrm2}^{'}\cos(\emph k_{\emph x}\emph a)\sin(\emph k_{\emph y}\emph a)\\
+\textrm2\textrm{i}\emph t_{\textrm{12}}^{''}\sin(\textrm2\emph k_{\emph y}\emph a)+\textrm4\textrm{i}\emph t_{\textrm1\textrm2}^{'''}\cos(\textrm2\emph k_{\emph x}\emph a)\sin(\emph k_{\emph y}\emph a),\\

\emph h_{\textrm2\textrm2}=\epsilon_{\textrm2}+\textrm2\emph t_{\textrm2\textrm2}\cos(\emph k_{\emph x}\emph a)+\textrm2\emph t_{\textrm3\textrm3}\cos(\emph k_{\emph y}\emph a)+\textrm4\emph t_{\textrm2\textrm2}^{'}\cos(\emph k_{\emph x}\emph a)\cos(\emph k_{\emph y}\emph a)\\
+\textrm2\emph t_{\textrm2\textrm2}^{''}\cos(\textrm2\emph k_{\emph x}\emph a)+\textrm2\emph t_{\textrm3\textrm3}^{''}\cos(\textrm2\emph k_{\emph y}\emph a)+\textrm4\emph t_{\textrm2\textrm2}^{'''}\cos(\textrm2\emph k_{\emph x}\emph a)\cos(\emph k_{\emph y}\emph a)+\textrm4\emph t_{\textrm3\textrm3}^{'''}\cos(\emph k_{\emph x}\emph a)\cos(\textrm2\emph k_{\emph y}\emph a),\\

\emph h_{\textrm2\textrm3}=-\textrm4\emph t_{\textrm2\textrm3}^{'}\sin(\emph k_{\emph x}\emph a)\sin(\emph k_{\emph y}\emph a)-\textrm4\emph t_{\textrm2\textrm3}^{'''}(\sin(\textrm2\emph k_{\emph x}\emph a)\sin(\emph k_{\emph y}\emph a)+\sin(\emph k_{\emph x}\emph a)\sin(\textrm2\emph k_{\emph y}\emph a)),\\

\emph h_{\textrm3\textrm3}=\epsilon_{\textrm2}+\textrm2\emph t_{\textrm3\textrm3}\cos(\emph k_{\emph x}\emph a)+\textrm2t_{\textrm2\textrm2}\cos(\emph k_{\emph y}\emph a)+\textrm4\emph t_{\textrm2\textrm2}^{'}\cos(\emph k_{\emph x}\emph a)\cos(\emph k_{\emph y}\emph a)\\
+\textrm2\emph t_{\textrm3\textrm3}^{''}\cos(\textrm2\emph k_{\emph x}\emph a)+\textrm2\emph t_{\textrm2\textrm2}^{''}\cos(\textrm2\emph k_{\emph y}\emph a)+\textrm4\emph t_{\textrm3\textrm3}^{'''}\cos(\textrm2\emph k_{\emph x}\emph a)\cos(\emph k_{\emph y}\emph a)+\textrm4\emph t_{\textrm2\textrm2}^{'''}\cos(\emph k_{\emph x}\emph a)\cos(\textrm2\emph k_{\emph y}\emph a),\\

\emph h_{\textrm4\textrm4}=\epsilon_{\textrm3}+\textrm2\emph t_{\textrm4\textrm4}[\cos(\emph k_{\emph x}\emph a)+\cos(\emph k_{\emph y}\emph a)]+\textrm4\emph t_{\textrm1\textrm1}^{'}\cos(\emph k_{\emph x}\emph a)\cos(\emph k_{\emph y}\emph a)\\
+\textrm2\emph t_{\textrm4\textrm4}^{''}[\cos(\textrm2\emph k_{\emph x}\emph a)+\cos(\textrm2\emph k_{\emph y}\emph a)]+\textrm4\emph t_{\textrm4\textrm4}^{'''}[\cos(\textrm2\emph k_{\emph x}\emph a)\cos(\emph k_{\emph y}\emph a)+\cos(\emph k_{\emph x}\emph a)\cos(\textrm2\emph k_{\emph y}\emph a)],\\

\emph h_{\textrm4\textrm5}=\textrm2\textrm{i}\emph t_{\textrm{45}}\sin(\emph k_{\emph x}\emph a)+\textrm4\textrm{i}\emph t_{\textrm4\textrm5}^{'}\sin(\emph k_{\emph x}\emph a)\cos(\emph k_{\emph y}\emph a)\\

\emph h_{\textrm4\textrm6}=\textrm2\textrm{i}\emph t_{\textrm{45}}\sin(\emph k_{\emph y}\emph a)+\textrm4\textrm{i}\emph t_{\textrm4\textrm5}^{'}\cos(\emph k_{\emph x}\emph a)\sin(\emph k_{\emph y}\emph a)\\

\emph h_{\textrm5\textrm5}=\epsilon_{\textrm4}+\textrm2\emph t_{\textrm5\textrm5}\cos(\emph k_{\emph x}\emph a)+\textrm2\emph t_{\textrm6\textrm6}\cos(\emph k_{\emph y}\emph a)+\textrm4\emph t_{\textrm5\textrm5}^{'}\cos(\emph k_{\emph x}\emph a)\cos(\emph k_{\emph y}\emph a)\\
+\textrm2\emph t_{\textrm5\textrm5}^{''}\cos(\textrm2\emph k_{\emph x}\emph a)+\textrm2\emph t_{\textrm6\textrm6}^{''}\cos(\textrm2\emph k_{\emph y}\emph a)+\textrm4\emph t_{\textrm5\textrm5}^{'''}\cos(\textrm2\emph k_{\emph x}\emph a)\cos(\emph k_{\emph y}\emph a)+\textrm4\emph t_{\textrm6\textrm6}^{'''}\cos(\emph k_{\emph x}\emph a)\cos(\textrm2\emph k_{\emph y}\emph a),\\

\emph h_{\textrm5\textrm6}=-\textrm4\emph t_{\textrm5\textrm6}^{'}\sin(\emph k_{\emph x}\emph a)\sin(\emph k_{\emph y}\emph a)-\textrm4\emph t_{\textrm5\textrm6}^{'''}(\sin(\textrm2\emph k_{\emph x}\emph a)\sin(\emph k_{\emph y}\emph a)+\sin(\emph k_{\emph x}\emph a)\sin(\textrm2\emph k_{\emph y}\emph a)),\\

\emph h_{\textrm6\textrm6}=\epsilon_{\textrm4}+\textrm2\emph t_{\textrm6\textrm6}\cos(\emph k_{\emph x}\emph a)+\textrm2t_{\textrm5\textrm5}\cos(\emph k_{\emph y}\emph a)+\textrm4\emph t_{\textrm5\textrm5}^{'}\cos(\emph k_{\emph x}\emph a)\cos(\emph k_{\emph y}\emph a)\\
+\textrm2\emph t_{\textrm6\textrm6}^{''}\cos(\textrm2\emph k_{\emph x}\emph a)+\textrm2\emph t_{\textrm5\textrm5}^{''}\cos(\textrm2\emph k_{\emph y}\emph a)+\textrm4\emph t_{\textrm6\textrm6}^{'''}\cos(\textrm2\emph k_{\emph x}\emph a)\cos(\emph k_{\emph y}\emph a)+\textrm4\emph t_{\textrm5\textrm5}^{'''}\cos(\emph k_{\emph x}\emph a)\cos(\textrm2\emph k_{\emph y}\emph a),
\end{array}
\end{equation}
and $\epsilon s$ are on-site energy, $\emph{ts}$ are hopping energy. The fitting parameters are listed in Table S1.
\begin{table*}[ptb]
	\caption{Fitting parameters in the six-band TBM of monolayer $\beta$-Bi$_{\textrm2}$Pd. $\epsilon$s are the on-site energy and ts are hopping energy. All energy parameters are in units of eV.}
	\begin{ruledtabular}
		\begin{tabular}{ccccccccccccccccc}
			$\epsilon_{\textrm1}$&$t_{\textrm{11}}$&$t_{\textrm{12}}$&$\epsilon_{\textrm2}$&$t_{\textrm{22}}$&$t_{\textrm{33}}$&$t_{\textrm{11}}^{'}$&$t_{\textrm{12}}^{'}$&$t_{\textrm{22}}^{'}$&$t_{\textrm{23}}^{'}$\\
			0.94&-0.55&0.06&0.78&1.86&-0.56&0.23&-0.25&0.24&0.31\\
			\hline
			$t_{\textrm{11}}^{''}$&$t_{\textrm{12}}^{''}$&$t_{\textrm{22}}^{''}$&$t_{\textrm{33}}^{''}$&$t_{\textrm{11}}^{'''}$&$t_{\textrm{12}}^{'''}$&$t_{\textrm{22}}^{'''}$&$t_{\textrm{23}}^{'''}$&$t_{\textrm{33}}^{'''}$\\
			0.07&0.04&0.18&0.05&0.04&0.03&0.02&0.02&-0.02\\
			\hline
			$\epsilon_{\textrm3}$&$t_{\textrm{44}}$&$t_{\textrm{45}}$&$\epsilon_{\textrm4}$&$t_{\textrm{55}}$&$t_{\textrm{66}}$&$t_{\textrm{44}}^{'}$&$t_{\textrm{45}}^{'}$&$t_{\textrm{55}}^{'}$&$t_{\textrm{56}}^{'}$\\
			-1.98&-0.75&-0.52&-0.34&1.94&-0.60&0.23&0.07&-0.18&-0.01\\
			\hline
			$t_{\textrm{44}}^{''}$&$t_{\textrm{55}}^{''}$&$t_{\textrm{66}}^{''}$&$t_{\textrm{44}}^{'''}$&$t_{\textrm{55}}^{'''}$&$t_{\textrm{56}}^{'''}$&$t_{\textrm{66}}^{'''}$\\
			0.07&0.18&-0.60&-0.04&-0.02&-0.02&0.02
		\end{tabular}
	\end{ruledtabular}
\end{table*}

\begin{figure}
	\centering
	\includegraphics[width=14cm]{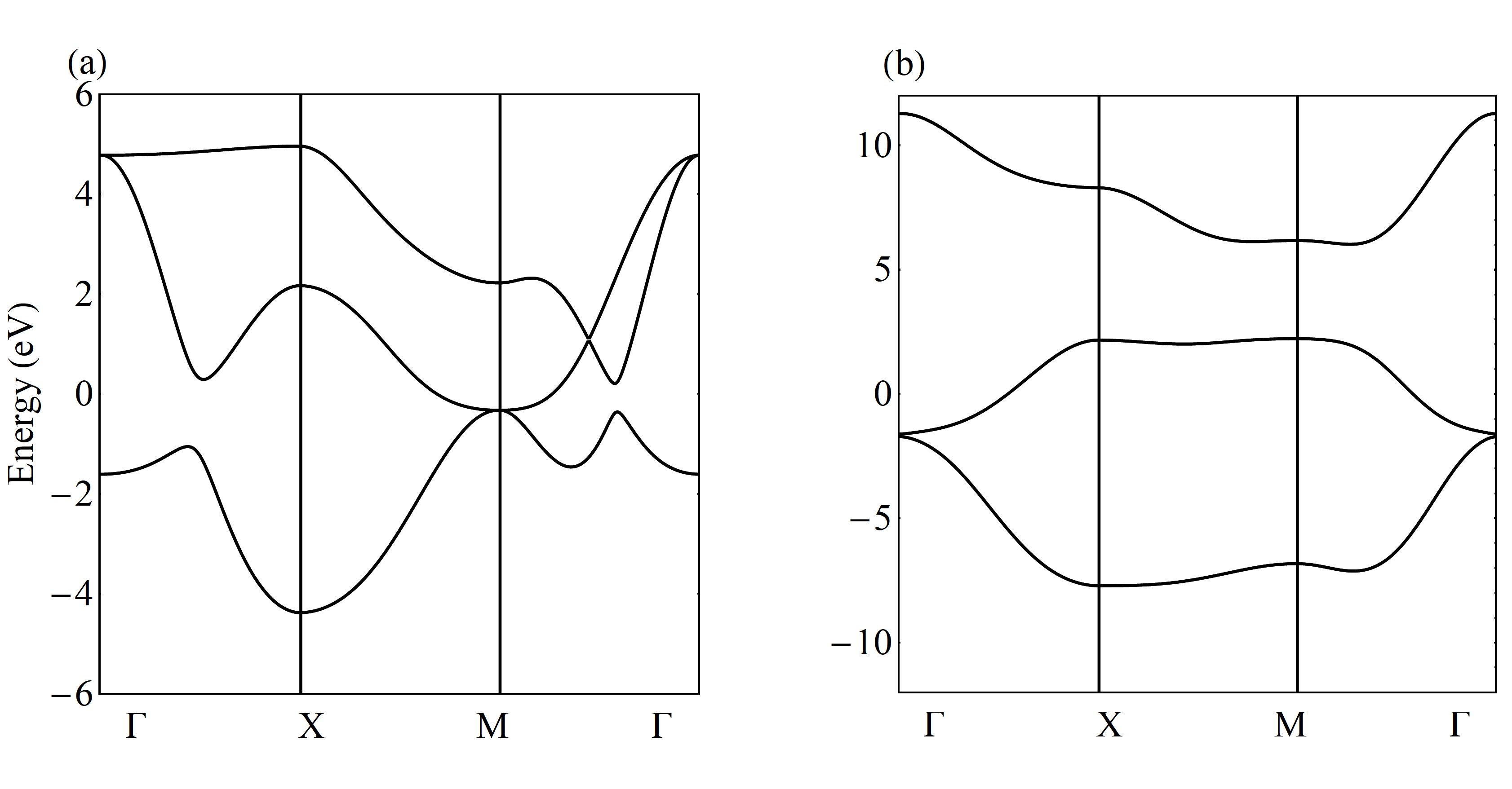}
	\caption{Energy spectrum of three-band TBM $\emph H_{\textrm 1}$ when $\lambda_{\textrm0}=$ 0 eV (a) and 6.39 eV (b).}\label{fig1}
\end{figure}
\section{\label{sec:level3}Spin-orbital coupling}
Owing to the heavy halogen family Bi, we consider the SOC effect under the basis 1: \{$|\textrm{Bi}_{\emph z}^{-}\rangle,|\textrm{Bi}_{\emph x}^{+}\rangle,|\textrm{Bi}_{\emph y}^{+}\rangle$\} and basis 2: \{$|\textrm{Bi}_{\emph z}^{+}\rangle,|\textrm{Bi}_{\emph x}^{-}\rangle,|\textrm{Bi}_{\emph y}^{-}\rangle$\},
\begin{equation}
\emph H_{\emph{SOC}}=\frac{\emph{dV}}{\emph{dr}}\frac{(\textbf{r}_{\emph i}\times \textbf{p}_{\emph i})\cdot \textbf{S}_{\emph i}}{\textrm2\emph r(\emph{mc})^{\textrm2}}=\textrm2\lambda \textbf{L}_{\emph i} \cdot \textbf{S}_{\emph i},
\end{equation}
where 
\begin{equation}
\emph L_{\emph x}=\sigma_{x}\otimes\left(\begin{array}{rrr}{\text0} & {\text1} & {\text0}\\ {\text1} & {\text0} & {\text0}\\ {\text0} & {\text0} & {\text0}\end{array}\right), 
\quad \emph L_{\emph y}=\sigma_{x}\otimes\left(\begin{array}{rrr}{\text0} & {\text0} & {\text1} \\ {\text0} & {\text0} & {\text0} \\ {\text1} & {\text0} & {\text0}\end{array}\right), 
\quad \emph L_{\emph z}=\sigma_{0}\otimes\left(\begin{array}{rrr}{\text0} & {\text0} & {\text0}\\ {\text0} & {\text0} & {-\imath}\\ {\text0} & {\imath} & {0}\end{array}\right).
\end{equation}
In our case, Rashba SOC is suppressed, i.e. $\emph L_{\emph x,y}$ is left out. So the full Hamiltonian is given by:
\begin{equation}
\emph H_{\emph S}(\textbf k)=\left[\begin{array}{lll}
\emph H_{\emph S\textrm1} & \\&\emph H_{\emph S\textrm2}
\end{array}\right]=\emph H_{\textrm0}(\textbf k)+\emph H_{\emph{SOC}}=\emph H_{\textrm0}(\textbf k)+2\lambda\emph L_{\emph z}\emph s_{\emph z},
\end{equation}
where $\emph s_{\emph z}$ is the z component of Pauli matrix.
\section{\label{sec:level3}Continuum model}
By the downfolding technique introduced before, the continuum Hamiltonian expanding at the M point for $\emph H_{\emph S\textrm1}$ to the leading order reads
\begin{equation}
\emph H_{\emph M}=\emph m_{\textrm0}\sigma_{\textrm0}+\emph m_{\textrm3}\emph k_{x}\emph k_{y}\sigma_{\emph x}+\lambda\sigma_{\emph y}+\frac{\emph m_{\textrm1}-\emph m_{\textrm2}}{\textrm2}(\emph k_{\emph x}^{\textrm2}-\emph k_{\emph y}^{\textrm2})\sigma_{\emph z},
\end{equation}
where $\sigma$ is the Pauli matrix on orbital space, $\emph m_{\textrm0}=\epsilon_{\textrm2}-\textrm2\left(\emph t_{\textrm{22}}+\emph t_{\textrm{33}}\right)+\textrm4 \emph t_{\textrm{22}}^{\prime}=$-0.86 just shifts energy, $\emph m_{\textrm1}=\left(\emph t_{\textrm{22}}-\textrm2\emph t_{\textrm{22}}^{\prime}\right)+\frac{\left(\textrm2\emph t_{\textrm{12}}-\textrm4\emph t_{\textrm{12}}^{\prime}\right)^{\textrm2}}{-\epsilon_{\textrm1}+\textrm4\emph t_{\text{11}} - \textrm4\emph t_{\text{11}}^{'}}$ $=$ 1.07, $\emph m_{\textrm2}=\emph t_{\textrm{33}}-\textrm2\emph t_{\textrm{22}}^{\prime}$ $=$-1.04, and $\emph m_{\textrm3}=-\textrm4 \emph t_{\textrm{23}}+\frac{\left(\textrm2\emph t_{\textrm{12}}-\textrm4 \emph t_{\textrm{12}}^{\prime}\right)^{\textrm2}}{-\epsilon_{\textrm1}+\textrm4\emph t_{\text{11}} - \textrm4\emph t_{\text{11}}^{'}}$ $=$-1.55. Here, we regard E = 0 eV and $\lambda$ as a small quantity and neglect the $[\frac{\emph m_{1}+\emph m_{2}}{2}(\emph k_{\emph x}^{2}+\emph k_{\emph y}^{2})]\sigma_{0}$ term which is trivial and negligible compared with $\emph m_{\textrm0}$. At the same time, we only take one block Hamiltonian in the spin subspace because the full Hamiltonian $\emph H_{\emph S\textrm1}(\textbf{k})$ is block diagonal and the two blocks are equivalent. Clearly, band structure around the DP disperses quadratic.

From Fig. S1 (b), one can see that energy spectrum closes at another TPTP $\Gamma$ point when $\lambda_{\text0}=$ 6.39 eV. It seems that the perturbation theory is invalid due to the quite large value of $\lambda$. However, we make a unitary transformation:
\begin{equation}
\emph U=\left(\begin{array}{lll}
1& 0&0\\
0 & -1&-\imath\\
0&1&-\imath
\end{array}\right)
\end{equation}
for Hamiltonian $\emph{UHU}^{\dagger}$ to simplify the problem. Then, we can get the continuum model at $\Gamma$ point as follow (upto the lowest terms of $\textbf k$):
\begin{equation}
\begin{array}{cc}
\emph H_{\Gamma}=\left(\begin{array}{cc}
\emph m_{\textrm1} & -\textrm i \gamma_{\textrm1}\emph k_{\emph x}-\gamma_{\textrm1}\emph k_{\emph y} \\
\textrm i \gamma_{\textrm2}\emph k_{\emph x}-\gamma_{\textrm1}\emph k_{\emph y}& \emph m_{\textrm2}-\lambda_{\textrm0}-\eta|\delta\lambda|
\end{array}\right)\\
=\gamma_{\textrm0}\sigma_{\textrm0}-\gamma_{\textrm1}\emph k_{\emph y}\sigma_{\emph x}+\gamma_{\textrm1}\emph k_{\emph x}\sigma_{\emph y}+\gamma_{\textrm2}\sigma_{\emph z},
\end{array}
\end{equation}
where $\emph m_{\textrm1}=\epsilon_{\textrm1}+\textrm4 \emph t_{\textrm{11}}+\textrm4 \emph t_{\textrm{11}}^{\prime}-\mu$ $=$ -0.34, $\emph m_{\textrm2}=\epsilon_{\textrm2}+\textrm2\left(\emph t_{\textrm{22}}+\emph t_{\textrm{33}}\right)+\textrm4 \emph t_{\textrm{22}}^{\prime}-\mu$ $=$4.34,
$\gamma_{\textrm0}=\emph m_{\textrm1}-\eta|\delta\lambda|/2$, $\gamma_{\textrm1}=(\textrm2\emph t_{\textrm{12}}+\textrm4\emph t_{\textrm{12}}^{'})/\sqrt{\textrm2}=$0.79, $\gamma_{\textrm2}=\eta|\delta\lambda|/2$. Here, the DP at the $\Gamma$ point disperses linear. 
\section{\label{sec:level3}Vortex structure}
\begin{figure}
	\centering
	\includegraphics[width=6cm]{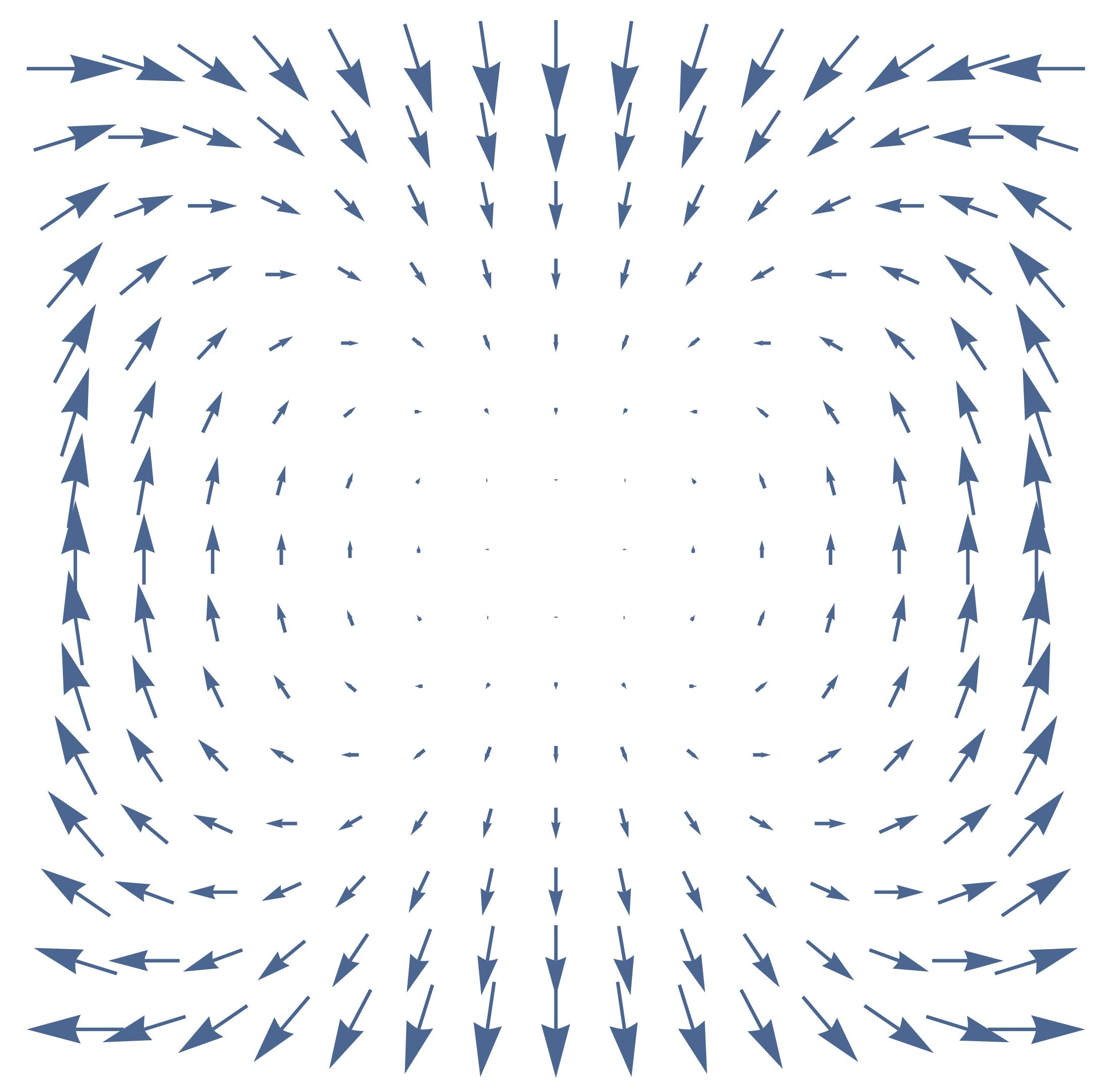}
	\caption{The planar vector $\textbf d$ in the momentum space near the DP.}\label{fig1}
\end{figure}
As shown in Fig. S2, the 2D planar vector defined by $\textbf d=\left[\emph m_{\textrm3} \emph k_{\emph x} \emph k_{\emph y}, \frac{\emph m_{\textrm1}-\emph m_{\textrm2}}{\textrm2}\left(\emph k_{\emph x}^{\textrm2}-\emph k_{\emph y}^{\textrm2}\right)\right]$ has a vortex structure at $\textbf k=\textrm0$, which indicates a band degeneracy point. The band degeneracy can be considered as a topological defect in the momentum space. 
\section{\label{sec:level3}$\mathcal{Z}_{\textrm2}$ index}
\begin{table}[ptb]
	\caption{Parities of valence and conduction bands at four time-reversal invariant momenta}
	\begin{ruledtabular}
		\begin{tabular}{l|cccccccccccccc}
			Time-reversal invariant momenta&$\Gamma$&$\emph X$&$\emph M$\\
			\hline
			Valence band&$-\textrm1$&$\textrm1$&$\textrm1$\\
			Conduction band&$\textrm1$&$-\textrm1$&$\textrm1$
		\end{tabular}
	\end{ruledtabular}
\end{table}
In 2D BZ, the $\mathcal{Z}_{\textrm2}$ number \cite{PhysRevLett.98.106803,fu2007topological} is defined by ($-$\textrm1)$^{\mathcal{Z}_{\textrm2}}$=$\prod^{\textrm4}_{\emph i=\textrm1}\delta(\emph P_{\emph i})$, where $\delta(\emph P_{\emph i})$ is the products of parity eigenvalues of the occupied bands below the band gap at the TRIM $\emph P_{\emph i}$. Parities of valence and conduction bands at four TRIM are shown in Table S2 which gives $\mathcal{Z}_{\textrm2}=$ 1.
\section{Topological Edge State}
\begin{figure}
	\centering
	\includegraphics[width=10cm]{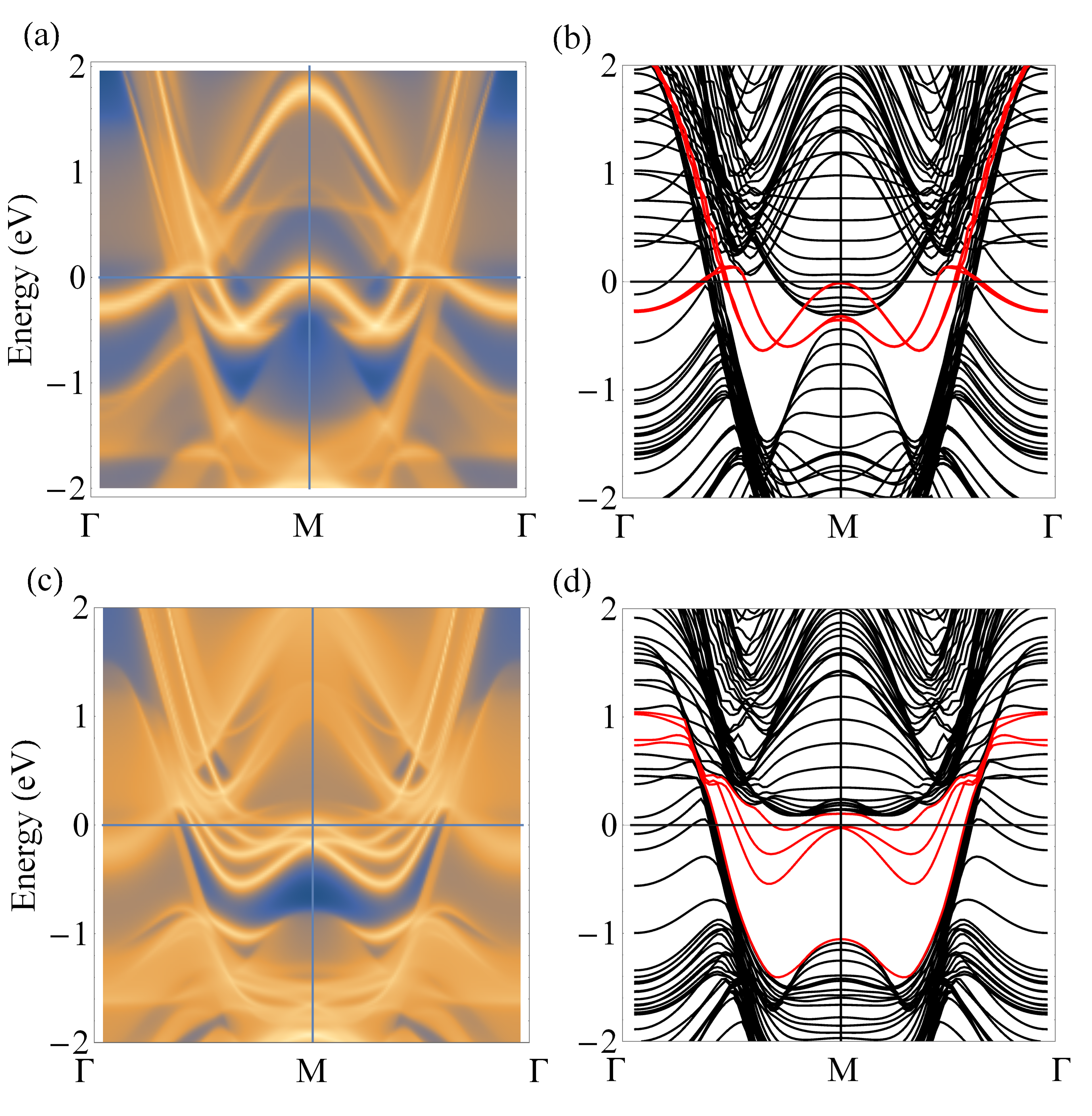}
	\caption{TESs without (a) (b) and with (c) (d) SOC. (a) (c) Bi terminated TESs which is calculated by WannierTools. (b) (d) TESs calculated according to our six-band TBM.}\label{fig1}
\end{figure}
To show the validity of our model, in Fig. S3, we show the Bi terminated TESs calculated by WannierTools as a control. One can see that our six-band TBM is effective and captures the main physics since the corresponding TESs that we are interested are quite similar to the results from first-princles calculations.
\section{\label{sec:level3}Evolution of Superconducting band gaps}
\begin{figure}
	\centering
	\includegraphics[width=15cm]{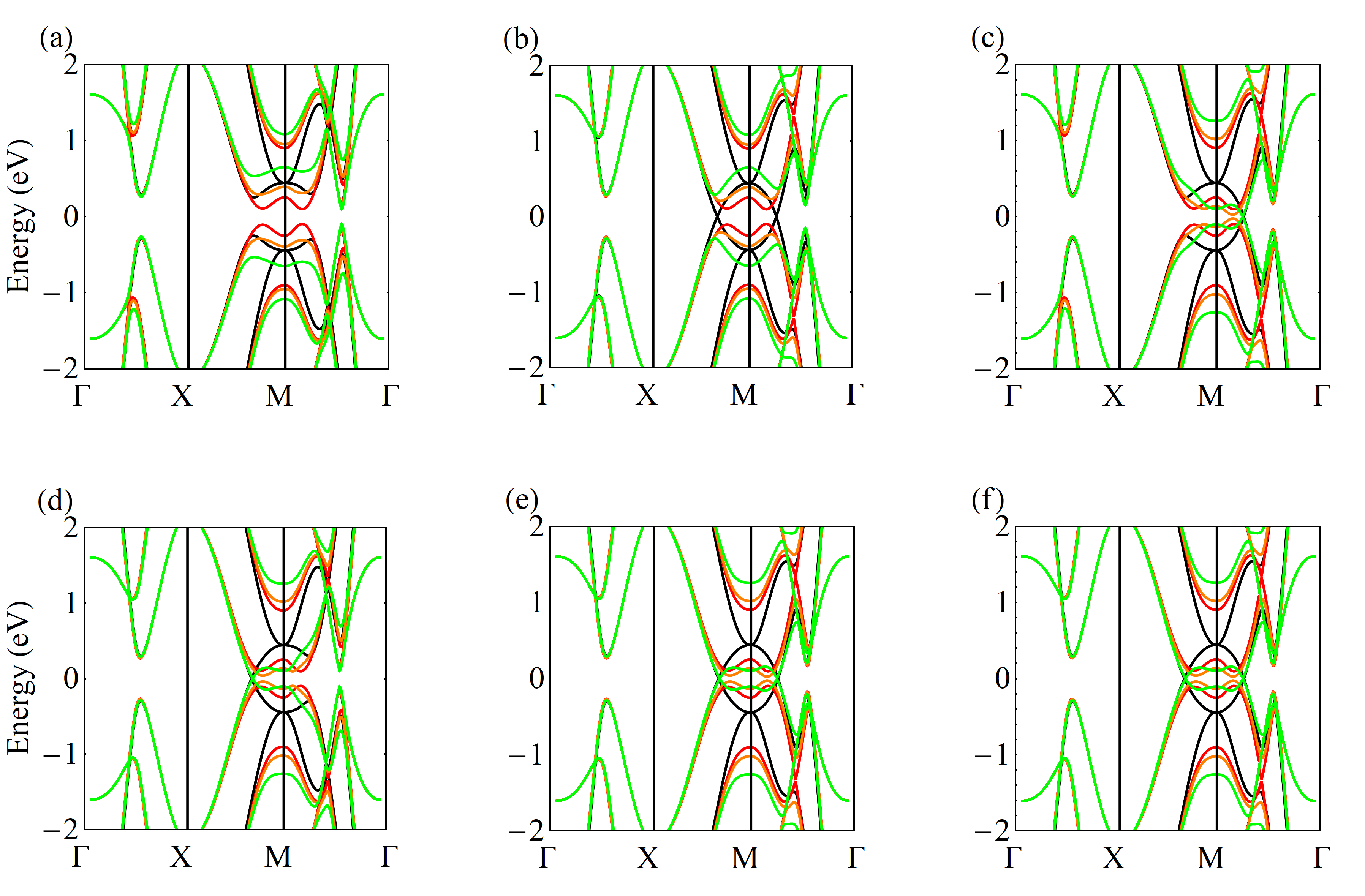}
	\caption{The superconducting band structures for different pairing potentials (a) $\Delta_{\emph s\textrm1}$, (b) $\Delta_{\emph s\textrm2}$, (c) $\Delta_{\emph s\textrm3}$, and (d) $\Delta_{\emph s\textrm4}$. Here, black/red/orange/green lines indicate $\lambda=0/0.578/0.578/0.578$ eV and $\Delta_{0}=0.3/0/0.3/0.6$ eV.}\label{fig1}
\end{figure}
\begin{figure}
	\centering
	\includegraphics[width=15cm]{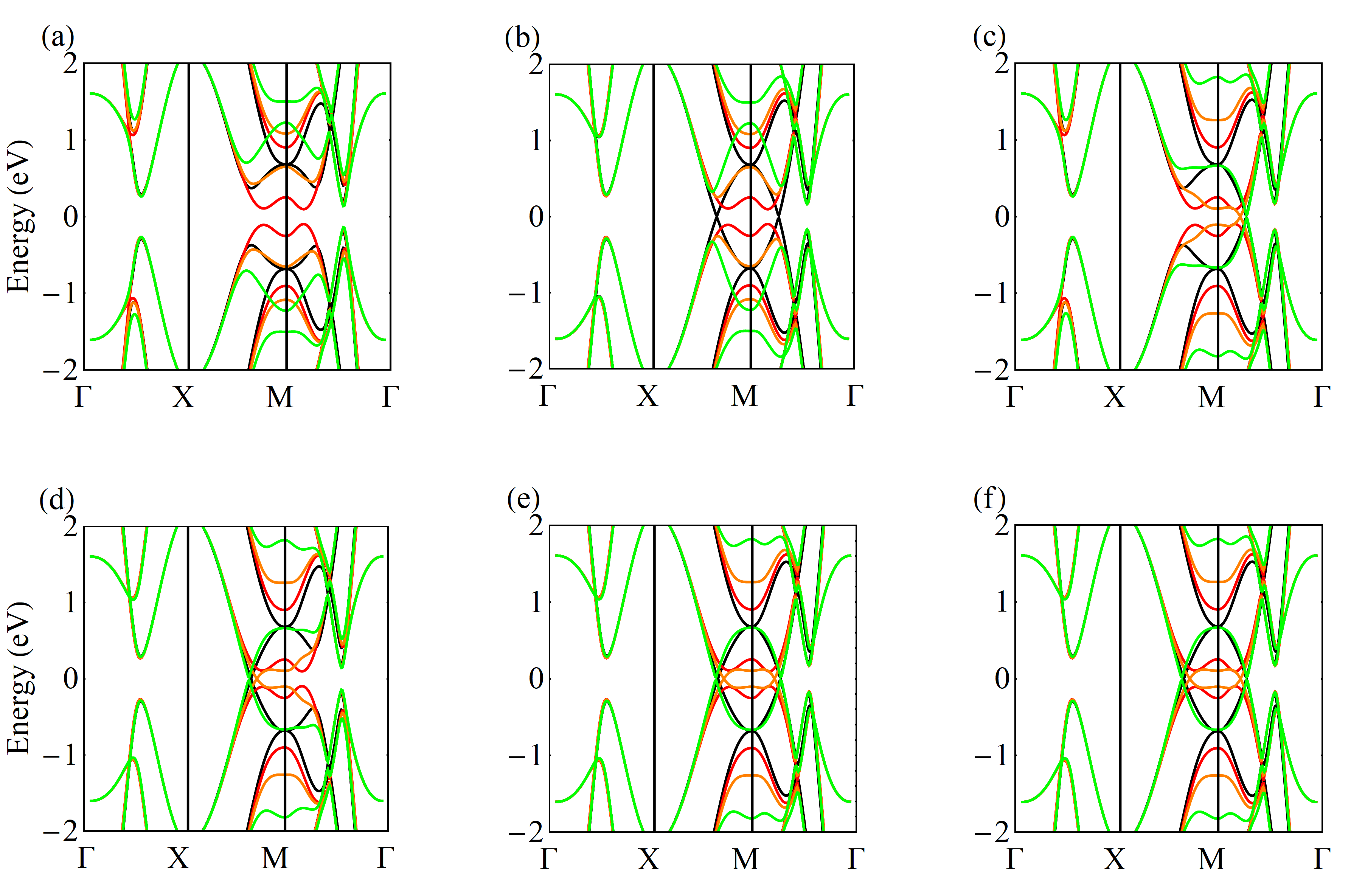}
	\caption{The superconducting band structures for different pairing potentials (a) $\Delta_{\emph s^{*}\textrm1}$, (b) $\Delta_{\emph s^{*}\textrm2}$, (c) $\Delta_{\emph s^{*}\textrm3}$, and (d) $\Delta_{\emph s^{*}\textrm4}$. Here, black/red/orange/green lines indicate $\lambda=0/0.578/0.578/0.578$ eV and $\Delta_{0}=0.3/0/0.3/0.6$ eV.}\label{fig1}
\end{figure}
\begin{figure}
	\centering
	\includegraphics[width=15cm]{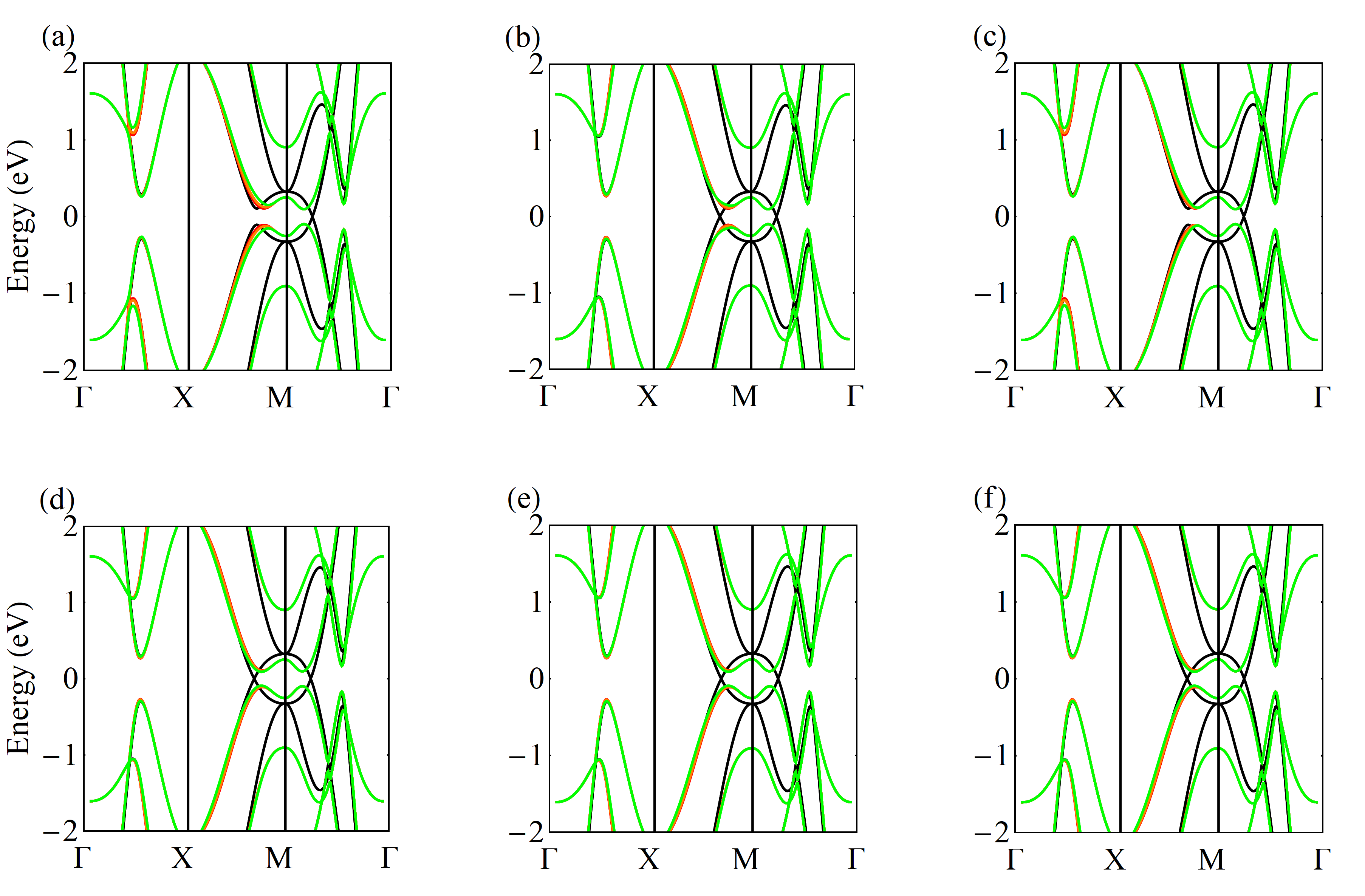}
	\caption{The superconducting band structures for different pairing potentials (a) $\Delta_{\emph d\textrm1}$, (b) $\Delta_{\emph d\textrm2}$, (c) $\Delta_{\emph d\textrm3}$, and (d) $\Delta_{\emph d\textrm4}$. Here, black/red/orange/green lines indicate $\lambda=0/0.578/0.578/0.578$ eV and $\Delta_{0}=0.3/0/0.3/0.6$ eV.}\label{fig1}
\end{figure}
\begin{figure}
	\centering
	\includegraphics[width=15cm]{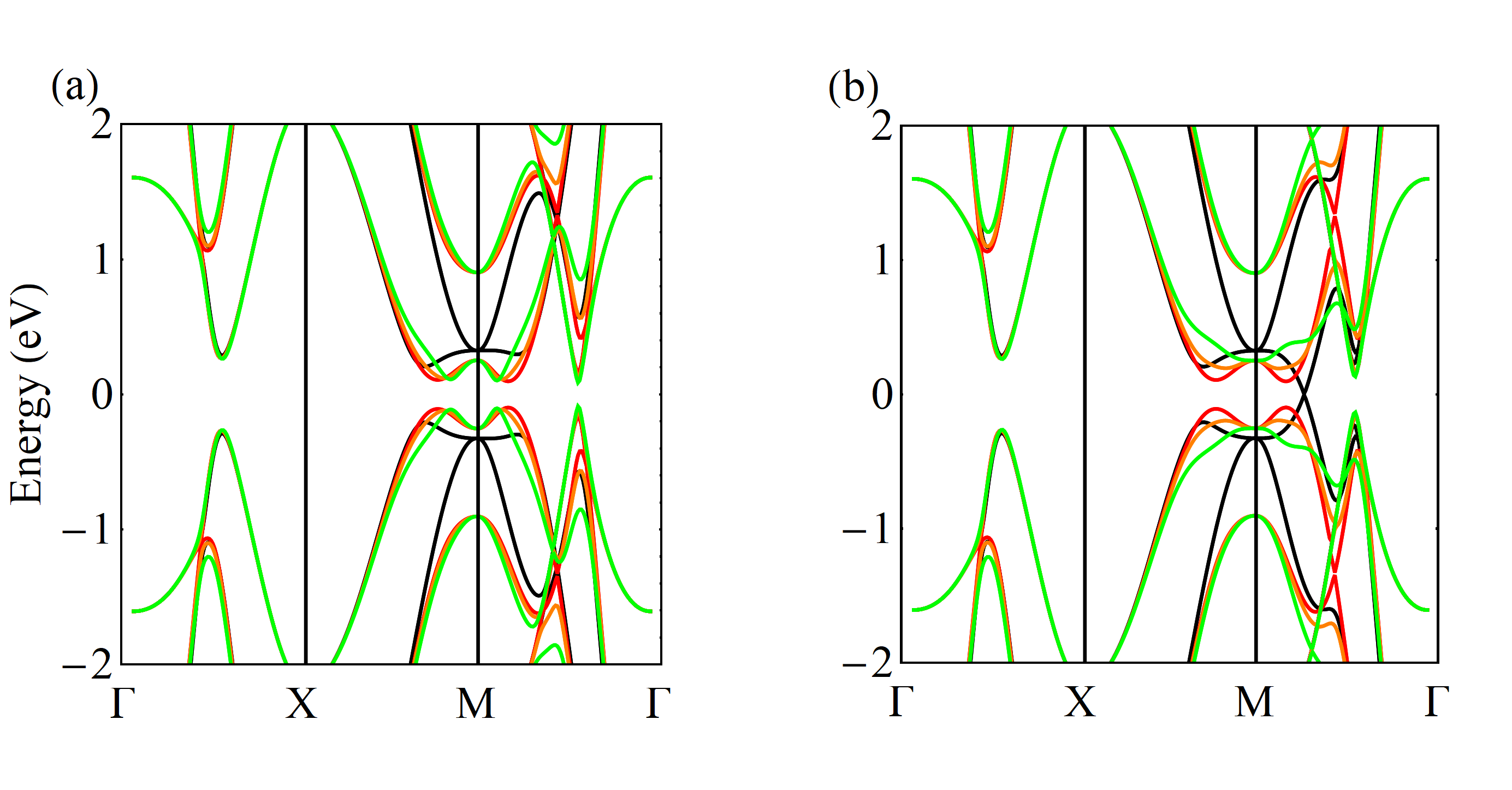}
	\caption{The superconducting band structures for different pairing potentials (a) $\Delta_{\emph p\textrm1}$, (b) $\Delta_{\emph p\textrm2}$. Here, black/red/orange/green lines indicate $\lambda=0/0.578/0.578/0.578$ eV and $\Delta_{0}=0.3/0/0.3/0.6$ eV.}\label{fig1}
\end{figure}
\begin{figure}
	\centering
	\includegraphics[width=8cm]{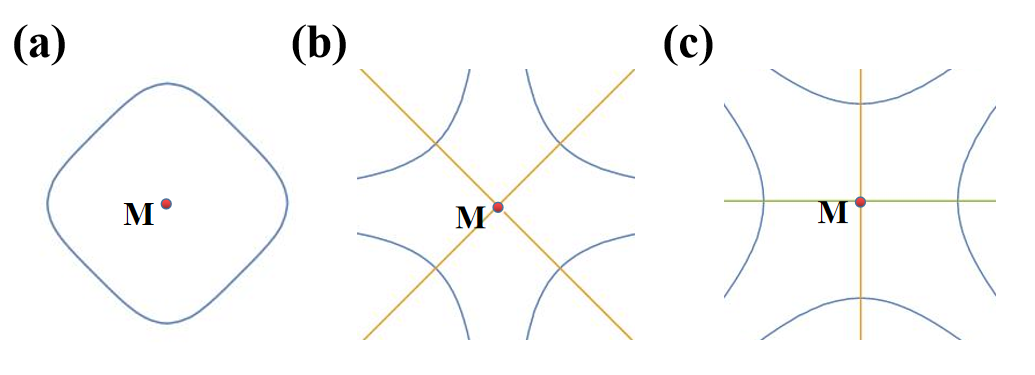}
	\caption{The diagrammatic sketch of $\textbf k$-equation around the M point for different pairings (a) $\Delta_{\emph s\textrm2/5}$, (b) $\Delta_{\emph s\textrm3}$ (c) $\Delta_{\emph s\textrm4}$.}\label{fig1}
\end{figure}
In Figs. S4-S7, we show the superconducting band structures for different pairing potentials.
The evolution of superconducting band gaps around the M point can be explain qualitatively by effectvie BdG $\textbf k\cdot\textbf p$ model at M point. We give the energy spectrum around the M point of $\Delta_{\emph s\textrm1}$ pairing potential first:
\begin{equation}
\begin{aligned}
\emph E^{\textrm2}=\Delta_{\textrm0}^{\textrm2}+\left[\frac{(\emph m_{\textrm1}-\emph m_{\textrm2})}{\textrm2}\left(\emph k_{\emph x}^{\textrm2}-\emph k_{\emph y}^{\textrm2}\right)\right]^{\textrm2}+(\emph m_{\textrm3}k_{\emph x}\emph k_{\emph y})^{\textrm2}+\lambda^{\textrm2}.
\end{aligned}
\end{equation}
It gives full superconducting gap only if the gap function $\Delta_{\textrm0}$ or $\lambda$ is nonzero. For $\Delta_{\emph s\textrm2}$ pairing potential, the eigenvalue is given by
\begin{equation}
\begin{aligned}
\emph E^{\textrm2}=\Delta_{\textrm0}^{\textrm2}+\left[\frac{(\emph m_{\textrm1}-\emph m_{\textrm2})}{\textrm2}\left(\emph k_{\emph x}^{\textrm2}-\emph k_{\emph y}^{\textrm2}\right)\right]^{\textrm2}+(\emph m_{\textrm3}\emph k_{\emph x}\emph k_{\emph y})^{\textrm2}+\lambda^{\textrm2}\pm \textrm2 \sqrt{\Delta_{\textrm0}^{\textrm2}\left\{\left[\frac{(\emph m_{\textrm1}-\emph m_{\textrm2})}{\textrm2}\left(\emph k_{\emph x}^{\textrm2}-\emph k_{\emph y}^{\textrm2}\right)\right]^{\textrm2}+(\emph m_{\textrm3}\emph k_{\emph x}\emph k_{\emph y})^{\textrm2}\right\}}.
\end{aligned}
\end{equation}
One can see when $\lambda$ is zero but $\Delta_{\textrm0}$ is nonzero, a Fermi loop around the M point is obtained by solving such $\textbf k$-equation
\begin{equation}
[\frac{(\emph m_{\textrm1}-\emph m_{\textrm2})}{\textrm2}(\emph k_{\emph x}^{\textrm2}-\emph k_{\emph y}^{\textrm2})]^{\textrm2}+(\emph m_{\textrm3}\emph k_{\emph x}\emph k_{\emph y})^{\textrm2}=\Delta_{\textrm0}^{\textrm2}.
\end{equation}
The diagrammatic sketch is shown in Fig. S8 (a) in which the loop is actually a rounded square.
Once turning on SOC, we will get a full superconducting gap. Next the spectrum of channel $\emph B_{\emph s\textrm1\emph g}$ reads
\begin{equation}
\begin{aligned}
\emph E^{\textrm2}=\Delta_{\textrm0}^{\textrm2}+\left[\frac{(\emph m_{\textrm1}-\emph m_{\textrm2})}{\textrm2}\left(\emph k_{\emph x}^{\textrm2}-\emph k_{\emph y}^{\textrm2}\right)\right]^{\textrm2}+(\emph m_{\textrm3}\emph k_{\emph x}\emph k_{\emph y})^{\textrm2}+\lambda^{\textrm2}\pm \textrm2 \sqrt{\Delta_{\textrm0}^{\textrm2}\{\lambda^{\textrm2}+\left[\frac{(\emph m_{\textrm1}-\emph m_{\textrm2})}{\textrm2}\left(\emph k_{\emph x}^{\textrm2}-\emph k_{\emph y}^{\textrm2}\right)\right]^{\textrm2}\}},
\end{aligned}
\end{equation}
and the $\textbf k$-equation is
\begin{equation}
(\emph m_{\textrm3}k_{\emph x}k_{\emph y})^{\textrm2}=\textrm0,\quad\text{and}\quad\lambda^{\textrm2}+\left[\frac{(\emph m_{\textrm1}-\emph m_{\textrm2})}{\textrm2}\left(k_{\emph x}^{\textrm2}-k_{\emph y}^{\textrm2}\right)\right]^{\textrm2}=\Delta_{\textrm0}^{\textrm2}.
\end{equation}
There are only four DPs on $\Gamma$-M line that satisfy the condition Eq. (21) as shown in Fig. 4 (b).
If $|\lambda|$ increases from zero to $|\Delta_{\textrm0}|$, these nodes will encounter each other at the M point and then is gapped when $|\lambda|$ $>$ $|\Delta_{\textrm0}|$. For channel $\emph B_{\emph s\textrm2\emph g}$, the eigenvalue is given by
\begin{equation}
\begin{aligned}
\emph E^{\textrm2}=\Delta_{\textrm0}^{\textrm2}+\left[\frac{(\emph m_{\textrm1}-\emph m_{\textrm2})}{\textrm2}\left(\emph k_{\emph x}^{\textrm2}-\emph k_{\emph y}^{\textrm2}\right)\right]^{\textrm2}+(\emph m_{\textrm3}\emph k_{\emph x}\emph k_{\emph y})^{\textrm2}+\lambda^{\textrm2}\pm \textrm2 \sqrt{\Delta_{\textrm0}^{\textrm2}[\lambda^{\textrm2}+(\emph m_{\textrm3}\emph k_{\emph x}\emph k_{\emph y})^{\textrm2}]},
\end{aligned}
\end{equation}
and the $\textbf k$-equation is
\begin{equation}
[\frac{(\emph m_{\textrm1}-\emph m_{\textrm2})}{\textrm2}(\emph k_{\emph x}^{\textrm2}-\emph k_{\emph y}^{\textrm2})]^{\textrm2}=\textrm0,\quad\text{and}\quad\lambda^{\textrm2}+(\emph m_{\textrm3}\emph k_{\emph x}\emph k_{\emph y})^{\textrm2}=\Delta_{\textrm0}^{\textrm2}.
\end{equation}
It is similar to channel $\emph B_{\textrm1\emph g}$ but four DPs move to X-M line. The eigenvalues of last pairing $\Delta_{\textrm5}$ is the same as $\Delta_{\textrm2}$. The situation in s$^{*}$-wave is similar to s-wave. However, in d-wave pairings, there is always a DP along M-$\Gamma$ line due to the factor $\emph k_{\emph x}^{\text2}-\emph k_{\emph y}^{\text2}$.
\section{\label{sec:level3}Superconducting susceptibility}
\begin{figure}
	\centering
	\includegraphics[width=7cm]{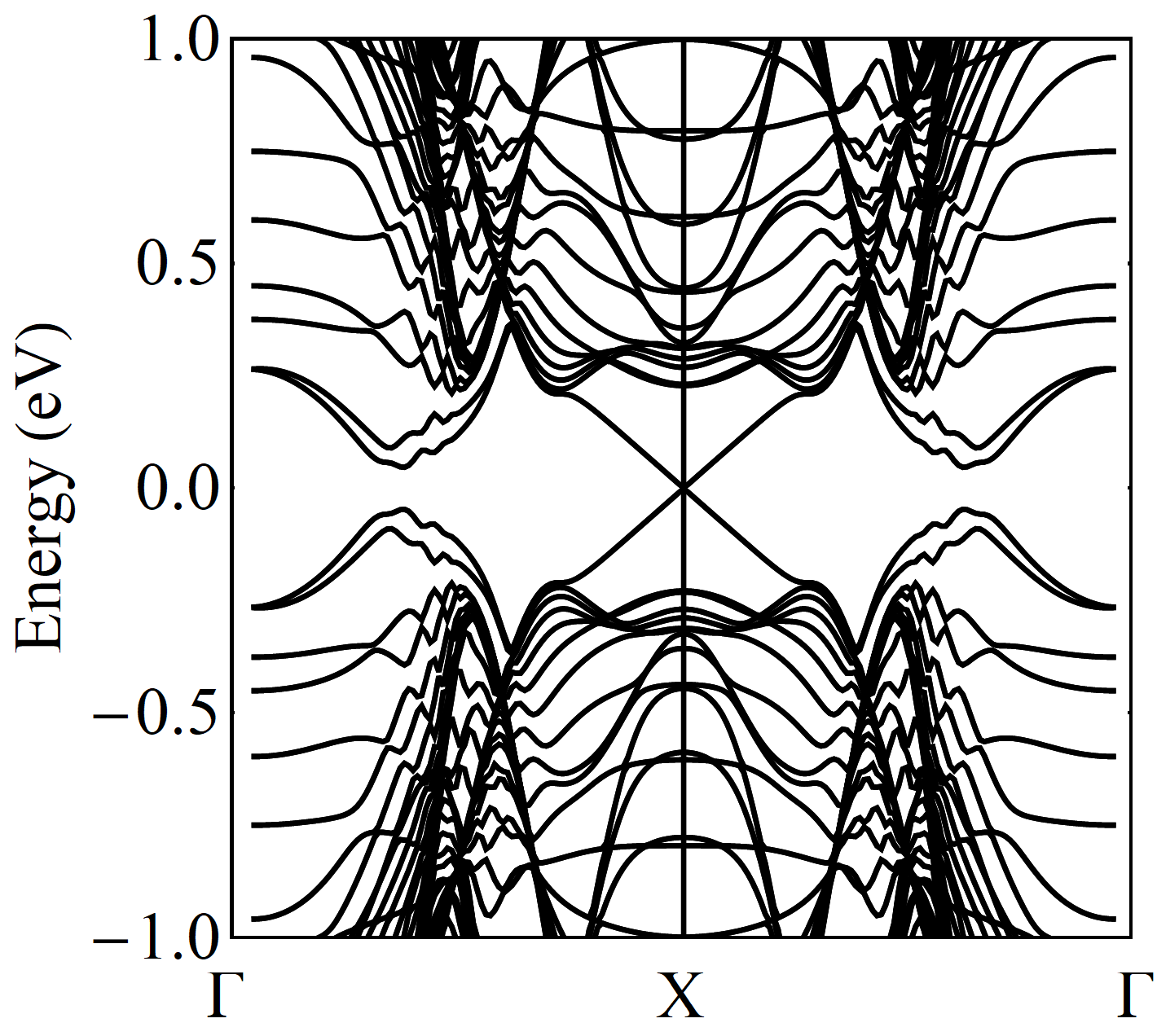}
	\caption{MZM for pairing $\Delta_{\emph p\text1}$ without SOC.}\label{fig1}
\end{figure}
To evaluate the linearized gap equations in each pairing potential, it is s essential to calculate superconductivity susceptibility. Since the Fermi surface is near the M point, we consider the continuum Hamiltonian at M point, which has been given in Eq. (14). Then, the finite temperature superconducting susceptibility $\chi_{\alpha\emph i}$ can be calculated by
\begin{equation}
\chi_{\alpha\emph i}=-\frac{\textrm1}{\beta} \sum_{\textrm i \omega_{\emph n}, \textbf k}\operatorname{Tr}\left[\Delta_{\alpha\emph i}^{\dagger}(\textbf k)\mathcal{G}_{\emph e}\left(\textbf k, \textrm i \omega_{\emph n}\right)\Delta_{\alpha\emph i}(\textbf k) \mathcal{G}_{\emph h}\left(\textbf k, \textrm i \omega_{\emph n}\right)\right],
\end{equation}
where $\mathcal{G}_{\emph e,\emph h}\left(\textbf k, \textrm i \omega_{\emph n}\right)=\left[\textrm i \omega_{\emph n} \mp \emph H_{\emph M}(\textbf k)\right]^{-\textrm1}$ are the relevant
standard electron and hole Matsubara Green functions. The pairings with different representations are independent of each other, and thus, we can compute the STT in each representation, separately.

1. s-wave ($\emph f_{\emph s}(\textbf k)=$ 1)

$\emph A_{\text1\emph g}$ representation:

We have two possible representation matrices for the $\emph A_{\text1\emph g}$ representation $\sigma_{\textrm0}s_{\textrm0}$ and $\sigma_{\emph y}\emph s_{\emph z}$. The superconductivity susceptibility is given by
$\chi_{\emph s,\emph A_{\text1\emph g},\text{11}}=\text4\chi_{\text0}$, $\chi_{\emph s,\emph A_{\text1\emph g},\text{12}}=\text4\chi_{\text0}\frac{\lambda}{\mu}$, $\chi_{\emph s,\emph A_{\text1\emph g},\text{12}}=\text4\chi_{\text0}\frac{\lambda^{\text2}}{\mu^{\text2}}$, where $\chi_{\emph A_{\text1\emph g}, \text{11}}$ is for $\sigma_{\textrm0}s_{\textrm0}$, $\chi_{\emph A_{\text1\emph g}, \text{22}}$ is for $\sigma_{\emph y}\emph s_{\emph z}$ and $\chi_{\emph A_{\text1\emph g}, \text{12}}$ describe the coupling between $\sigma_{\textrm0}s_{\textrm0}$ and $\sigma_{\emph y}\emph s_{\emph z}$, and $\chi_{\textrm0}=\emph g(\textrm0) \int \emph d \varepsilon \tanh \left(\frac{\varepsilon}{2 \emph k_{\emph B} \emph T}\right) / \varepsilon$
is the standard superconducting susceptibility, in which $\emph g(\textrm0)$ is the density of states at the Fermi level. So, we can obtain a set of linear equations
\begin{equation}
\begin{aligned}
&\Delta_{\emph s,\emph A_{\text1\emph g}, \text1}=U(\chi_{\emph s,\emph A_{\text1\emph g},\text{11}} \Delta_{\emph s,\emph A_{\text1\emph g}, \text1}+\chi_{\emph s,\emph A_{\text1\emph g},\text{12}} \Delta_{\emph s,\emph A_{\text1\emph g}, \text1}), \\
&\Delta_{\emph s,\emph A_{\text1\emph g}, \text1}=V(\chi_{\emph s,\emph A_{\text1\emph g},\text{12}} \Delta_{\emph s,\emph A_{\text1\emph g}, \text1}+\chi_{\emph s,\emph A_{\text1\emph g},\text{22}} \Delta_{\emph s,\emph A_{\text1\emph g}, \text1}).
\end{aligned}
\end{equation}
It can be viewed as an eigen-equation and  gives rise to the STT
\begin{equation}
\emph k_{\emph B} \emph T_{\emph c}=\frac{\textrm2 \emph e^{\gamma}}{\pi}\omega_{\emph D} \exp \left(-\frac{\textrm1}{\textrm2 \emph g(\textrm0) (\emph U+\emph V\frac{\lambda^{\text2}}{\mu^{\text2}})}\right).
\end{equation}
The corresponding eigen-state satisfies
\begin{equation}
\Delta_{\emph s,\emph A_{\text1\emph g}, \text2}=\frac{V\lambda}{U\mu}\Delta_{\emph s,\emph A_{\text1\emph g}, \text1}.
\end{equation}
In the strong SOC limit $\lambda\approx\mu$, the superconductivity susceptibilities of other representations are all zero. Thus, the most possible pairing is $\Delta_{\emph s,\emph A_{\text1\emph g}}$. For non-on-site pairing, rather than performing an accurate calculation of STT \cite{PhysRevLett.101.206404,PhysRevX.1.011009,Hu2012}, we qualitative evaluate the relative relation of STT in each channel. The most possible pairings are also representation $\emph A_{\text1\emph g}$ in s$^{*}$-wave and d-wave pairings.

2. p-wave [$\emph f_{\emph p}(\textbf k)=(\emph k_{\emph x},\emph k_{\emph y})$]

In the strong SOC limit, the superconductivity susceptibilities is given by $\chi_{\emph p,\emph A_{\text1\emph u},\text{11}}=0$, $\chi_{\emph p,\emph A_{\text1\emph u},\text{12}}=0$, and $\chi_{\emph p,\emph A_{\text1\emph u},\text{22}}=\text4\chi_{\emph p}\frac{\lambda^{\text2}}{\mu^{\text2}}\approx\text4\chi_{\emph p}$, where $\chi_{\emph p}=\emph g(\textrm0) \int \emph d \varepsilon |\emph f_{\emph p}(\textbf k)|^{\text2}\tanh \left(\frac{\varepsilon}{2 \emph k_{\emph B} \emph T}\right) / \varepsilon$.
\bibliographystyle{apsrev4-1}
\bibliography{bibl}